\documentclass{emulateapj} 

\usepackage{amsmath}
\usepackage{amsfonts}
\usepackage{amssymb}
\usepackage{bm}
\usepackage{graphicx}
\usepackage{times}

\begin{document}

\title{Dependence of solar wind power spectra on the direction of the local mean magnetic field}

\author{J. J. Podesta}
\affil{Space Science Center, University of New Hampshire, Durham, NH 03824}
\email{jpodesta@solar.stanford.edu}

\begin{abstract}
Wavelet analysis can be used to measure the power spectrum of solar wind fluctuations 
along a line in any direction $(\theta, \phi)$ with respect to the local mean magnetic field
$\bm B_0$.
This technique is applied 
to study solar wind turbulence in high-speed streams in the ecliptic plane near solar minimum
using magnetic field measurements with a cadence of eight vectors per second.  
The analysis of nine high-speed streams 
shows that the reduced spectrum of magnetic field fluctuations (trace power) 
is approximately azimuthally symmetric about $\bm B_0$ in both the inertial range and 
dissipation range; in the inertial range the spectra are characterized by a power-law exponent 
that changes continuously from $1.6\pm 0.1$ in the direction perpendicular to the mean field 
to $2.0\pm 0.1$ in the direction parallel to the mean field.
The large uncertainties suggest that the perpendicular power-law indices 
3/2 and 5/3 are both consistent with the data.
The results are similar to those found by Horbury et al. (2008) at high heliographic latitudes.
Comparisons between solar wind observations and the theories of strong incompressible MHD
turbulence developed by Goldreich \& Sridhar (1995) and Boldyrev (2006) are not
rigorously justified because these theories only apply to turbulence with vanishing
cross-helicity although the normalized cross-helicity of solar wind turbulence is not negligible.
Assuming these theories can be generalized in such a way that the 3D wavevector spectra 
have similar functional forms when the cross-helicity is nonzero, then for the interval 
of Ulysses data analyzed by 
Horbury et al. (2008) the ratio of the spectra perpendicular and parallel
to $\bm B_0$ is more consistent with the Goldreich \& Sridhar scaling 
$P_\perp/P_\parallel \propto \nu^{1/3}$ than with the Boldyrev scaling $\nu^{1/2}$.  
The analysis of high speed streams in the ecliptic plane does not yield a reliable 
measurement of this scaling law.  
The transition from a turbulent MHD-scale energy cascade to a kinetic Alfv\'en wave (KAW)
cascade occurs when $k_\perp \rho_i\simeq 1$ which coincides with the spectral break.
At slightly higher wavenumbers,
in the dissipation range, there is a peak in the power 
ratio with $P_\perp/P_\parallel \gg 1$.  The decay of this peak 
may be caused by the damping of KAWs which is predicted to occur near $k_\perp \rho_i\simeq 4$.
\end{abstract}


\section{Introduction}

In the presence of a strong mean magnetic field $\bm B_0$, MHD turbulence is 
spatially anisotropic: velocity and magnetic field fluctuations vary more rapidly in the direction
perpendicular to $\bm B_0$ than parallel to $\bm B_0$.  It is important to recognize that 
fluctuations at a given scale are most strongly affected by the {\it local} 
mean magnetic field at roughly 3 to 5 times that scale.  When this scale
dependence of the local mean magnetic field is taken into account, the anisotropy of the turbulence
becomes scale dependent because the direction of the local 
mean magnetic field changes with scale. Thus, anisotropy is a local
property of the turbulence that depends on both position and scale.
\medskip

In light of this, it is natural to expect wavelet analysis to be an effective 
tool for the study of anisotropy in plasma turbulence.  Wavelet analysis allows a
signal $f(t)$ to be decomposed into components that are localized in both 
time and frequency (or wavelet scale).  Therefore, unlike the Fourier transform, the wavelet 
transform is well adapted to the study of transient signals 
\citep{Daubechies:1992}.
Because of these properties, wavelet analysis can be used to study the 
power spectrum of turbulent fluctuations as a function of the direction of the local mean 
magnetic field as first shown by \citet{Horbury_Forman:2008} using solar wind data.   
\medskip

Horbury et al. (2008) studied a 30 day interval of 1 second magnetic field data acquired by
the Ulysses spacecraft at high heliographic latitudes during solar minimum conditions in 1995 
at a heliocentric distance of $\sim 1.4$ AU.
The high latitude flow of the solar wind at this time can be characterized as high speed, 
highly Alfv\'enic (large magnitude of the normalized cross-helicity), 
with the average magnetic field directed roughly radially outward at the largest scales.
By binning the square amplitude of the wavelet coefficients at a given scale 
according to the direction of the local mean magnetic field $\bm B_0$ at that scale 
they estimated the average power as a function of frequency and of the direction of $\bm B_0$.
They found that the power law index of inertial range fluctuations varies from
approximately 5/3 to approximately 2 as the angle $\theta$ between $\bm B_0$ and the
average flow direction decreases from $\pi/2$ to 0.  
\medskip

This was interpreted by Horbury et al. (2008) as evidence that the 3D wavevector spectrum of solar 
wind fluctuations has the form given in the \citet{Goldreich_Sridhar:1995}
theory of strong incompressible MHD turbulence.  This interpretation is somewhat
paradoxical, however, because the Goldreich \& Sridhar (1995) theory does not apply
to turbulence with non-vanishing cross-helicity as is usually found in the solar wind.
Nevertheless, the results are intriguing because they demonstrate that the power
law index of solar wind fluctuations varies with the angle $\theta$ between the
flow direction and the local mean magnetic field, at least for the
one data record studied by Horbury et al. (2008).  This has never been seen before.
In fact, based on earlier work by \citet{Sari_Valley:1976} 
and possibly others, it is
a common belief that the spectral exponent of solar wind fluctuations is independent of the 
angle $\theta$; a view confirmed in a recent study by Tessein et al. (2009).  However, the study
by Tessein et al. (2009) does not take into account the scale dependence of the
local mean magnetic field and this may explain, in part, why no dependence of the 
spectral exponents on $\theta$ was found in that study.
\medskip

Near solar minimum, the solar wind in the ecliptic plane is composed of recurring 
high-speed streams interspersed with lower speed wind.  These streams originate in 
equatorial coronal holes which extend from the polar regions of the sun to the equator  
\citep{Zirker:1977,Schwenn:2006} and have physical characteristics that 
are similar to the high-speed, high-latitude, solar wind found above polar coronal holes
near solar minimum.
Therefore, the variation in the power law index seen in high latitude Ulysses data
by Horbury et al. (2008) should also be present in high-speed streams in the ecliptic
plane around solar minimum.  The purpose of this study is to investigate this expectation
using magnetometer data from NASA's two Stereo spacecraft and to interpret the results using
anisotropic theories of incompressible MHD turbulence.  It should be emphasized that
the physical quantity measured in this study is the so called {\it reduced spectrum}
which is an integral of the 3D wavevector spectrum over a plane perpendicular 
to the mean magnetic field \citep{Matthaeus_Goldstein:1982, Podesta:2009}.


\section{Data selection}

NASA's two identically equiped Stereo spacecraft, Stereo Ahead (STA) and Stereo Behind (STB), 
were launched on 25 October 2006.  The spacecraft both follow an earth-like orbit with one
spacecraft moving progressively ahead of the earth and the other moving progressively 
behind the earth.  
This study uses data from the Stereo flux gate magnetometer \citep{Acuna:2008}
part of the IMPACT instrument \citep{Luhmann:2008}.
The data has a uniform cadence of 8 vectors per second and contains no
data gaps or missing data over the time intervals studied here.  
\medskip

\begin{deluxetable*}{cccccccccc}
\tablewidth{0pt}
\tablecaption{High-speed streams in the ecliptic plane near solar minimum}
\tablehead{
  \colhead{\#} &  \colhead{Year} &  \colhead{Begin} &  \colhead{End} &  \colhead{Days} &  
  \colhead{SC\tablenotemark{a}} &  
  \colhead{$\nu_1$ (Hz)\tablenotemark{b}} & \colhead{$\nu_2$ (Hz)\tablenotemark{b}} &
  \colhead{\begin{tabular}{c} Perpendicular \\ Exponent\tablenotemark{c} \end{tabular}} &
  \colhead{\begin{tabular}{c}  Exponent\\ --All data\tablenotemark{d} \end{tabular}}}
\startdata
1  & 2007 & 28 Apr 00:00 & 01 May 00:00  &   3   &  STB  &  $1\times 10^{-2}$  &  $2\times 10^{-1}$  &  $1.60\pm 0.07$  &  $1.62\pm 0.015$      \\
2  & 2007 & 25 May 00:00 & 28 May 02:39  &  3.11 &  STB  &  $4\times 10^{-3}$  &  $1\times 10^{-1}$  &  $1.55\pm 0.07$  &  $1.58\pm 0.015$   \\
3  & 2007 & 27 Aug 12:00 & 30 Aug 12:00  &   3   &  STB  &  $7\times 10^{-3}$  &  $3\times 10^{-1}$  &  $1.70\pm 0.07$  &  $1.63\pm 0.015$    \\
4  & 2007 & 15 Nov 00:00 & 18 Nov 00:00  &   3   &  STA  &  $7\times 10^{-3}$  &  $2\times 10^{-1}$  &  $1.55\pm 0.07$  &  $1.61\pm 0.015$     \\
5  & 2008 & 08 Jan 00:00 & 11 Jan 00:00  &   3   &  STA  &  $5\times 10^{-3}$  &  $2\times 10^{-1}$  &  $1.52\pm 0.07$  &  $1.57\pm 0.015$   \\
6  & 2008 & 13 Feb 00:00 & 18 Feb 00:00  &   5   &  STA  &  $7\times 10^{-3}$  &  $2\times 10^{-1}$  &  $1.65\pm 0.07$  &  $1.60\pm 0.015$     \\
7  & 2008 & 08 Mar 00:00 & 11 Mar 00:00  &   3   &  STB  &  $1\times 10^{-2}$  &  $2\times 10^{-1}$  &  $1.55\pm 0.07$  &  $1.61\pm 0.015$  \\
8  & 2008 & 04 Apr 00:00 & 08 Apr 00:00  &   4   &  STB  &  $5\times 10^{-3}$  &  $2\times 10^{-1}$  &  $1.51\pm 0.07$  &  $1.62\pm 0.015$    \\
9  & 2008 & 02 May 12:00 & 06 May 00:00  &  3.5  &  STB  &  $7\times 10^{-3}$  &  $2.5\times 10^{-1}$  &  $1.61\pm 0.07$  &  $1.63\pm 0.015$     
\enddata
\tablenotetext{a}{Spacecraft: STA $=$ Stereo A, STB $=$ Stereo B}
\tablenotetext{b}{The fits used to determine the power law exponents as a function of the angle $\theta$
were performed over the frequency interval $\nu_1<\nu<\nu_2$.}
\tablenotetext{c}{Power law index when $\theta\simeq 90$ degrees obtained from the results in Fig.\ 
\ref{exponents}.}
\tablenotetext{d}{Power law index over the frequency interval
$10^{-3}<\nu< 10^{-1}$ Hz obtained using all data for the indicated time interval.}
\end{deluxetable*}

\begin{figure*}
\begin{center}
\includegraphics[width=6in]{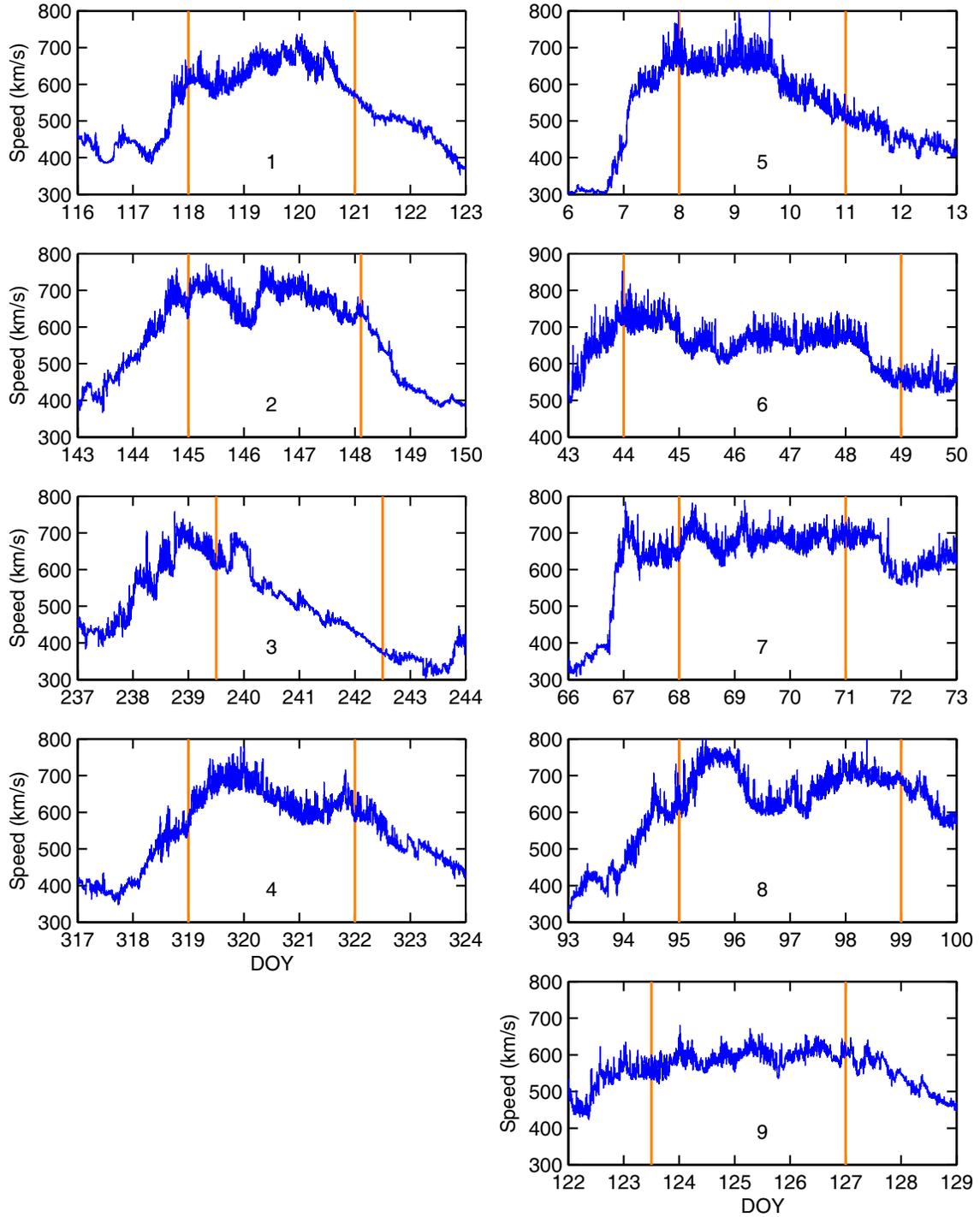}
\caption{\label{streams}%
Speed profiles obtained from the plasma instrument (PLASTIC) on Stereo.
The time intervals used in the wavelet analysis lie between the two vertical lines.
The labels 1 through 9 correspond to the time intervals in Table 1.
}
\end{center}
\end{figure*}

Between April 2007 and June 2008 the solar wind in the ecliptic plane consisted
a succession of high-speed streams of varying amplitudes interspersed with low-speed 
wind---typical conditions near solar minimum.  The data used in this study is 
restricted to high-speed streams; the leading and trailing edges of high-speed
streams were usually excluded.
High-speed streams of long duration with sustained high speeds were prefered over shorter 
lived more rapidly decaying streams because they provided larger contiguous samples of high
speed wind.  Streams were
identified using plasma data, proton velocity and density, from the PLASTIC instrument on 
board the two Stereo spacecraft \citep{Galvin:2008}.
The high-speed streams analyzed in this
study are displayed in Figure \ref{streams}
and their properties are listed in Table 1.  The length of each data record is typically 3 or 4 days.
\medskip

Near solar minimum, high speed streams are produced by equatorial coronal holes and are believed to be
similar to the high speed wind at high heliographic latitudes.  A single stream is usually embedded 
in a single magnetic sector of the interplanetary magnetic field and contains
no sector crossings.  For the nine intervals listed in
Table 1, this was checked using running averages of the magnetic field vector to create 2D
scatter plots of the average values of $(B_R,B_T)$ versus time.  Averaging windows
with durations of 20, 40, and 60 minutes were used and, for simplicity, the offset from the 
beginning of one averaging window to the beginning of the next window was half the window width.  
The data is assumed  to be contained in a single magnetic sector if the scatter plot remains 
in the same quadrant, either the second ($\bm B_0$ inward) or the fourth ($\bm B_0$ outward), 
for a majority of the time with no points or very few points in the
opposite quadrant (inward or outward).     
\medskip

The only other criteria used in the data selection is the requirement that power spectra
obtained by the wavelet analysis described below should form well defined power laws throughout
the inertial range and at all angles of interest.  In one instance, some of the spectra 
obtained using data from December 2007
showed unusual fluctuations and erratic points suggesting there may have been inadequate
sampling of the power in some angle bins or unusual statistical 
fluctuations in the time series, possibly caused by solar wind transients.
Therefore, this interval was discarded.

\section{Wavelet analysis technique}

\subsection{Continuous wavelet transform}

\indent\indent 
The continuous wavelet transform of a function $f(t)$ is defined by
\begin{equation}
F(s,t) = \int_{-\infty}^\infty |s|^{-1/2}\psi^* \bigg(\frac{\tau-t}{s}\bigg) f(\tau)\, d\tau,
\label{xfm}
\end{equation}
where $s$ is the wavelet scale, $t$ is the time, $\psi(t)$ is the mother wavelet, 
$\int_{-\infty}^\infty |\psi(t)|^2 \, dt = 1$, and the asterisk `$*$' denotes the complex conjugate.
The wavelet scale $s$ is related to the Fourier frequency as discussed below.   
The continuous wavelet transform allows any square integrable function $f(t)$ to be written as 
a continuous superposition of scaled and time-shifted wavelets by means of the 
inversion formula (Daubechies 1992)
\begin{equation}
f(t) = \frac{1}{C}\int_{-\infty}^\infty \int_{-\infty}^\infty |s|^{-1/2}\psi \bigg(\frac{t-\tau}{s}\bigg) 
F(s,\tau)\, \frac{ds \,d\tau}{s^2}
\label{inversion}
\end{equation}
provided 
\begin{equation}
C = \int_{-\infty}^\infty \frac{|\hat\psi(\omega)|^2}{|\omega|} \, d\omega <\infty,
\label{C}
\end{equation}
where
\begin{equation}
\hat\psi(\omega) = \int_{-\infty}^\infty\psi(t)  e^{-i\omega t}\, dt
\end{equation}
is the Fourier transform of $\psi(t)$.  
The distribution of energy in the $(s,t)$ plane (related to the time-frequency plane) is indicated 
by the relation
\begin{equation}
\int_{-\infty}^\infty |f(t)|^2\, dt = \frac{1}{C}\int_{-\infty}^\infty \int_{-\infty}^\infty 
|F(s,t)|^2\, \frac{ds \,dt}{s^2},
\label{Parseval}
\end{equation}
which is similar to Parseval's formula.
\medskip

In the special case where $f(t)$ is real valued and
$\hat\psi(\omega)$ vanishes for $\omega<0$, the inversion formula takes the simplified form
\begin{equation}
f(t) = \frac{2}{C}\int_{0}^\infty \frac{ds}{s^2}\int_{-\infty}^\infty d\tau \;
\mbox{Re}\, \bigg[s^{-1/2}\psi \bigg(\frac{t-\tau}{s}\bigg) F(s,\tau)\bigg],
\label{inv}
\end{equation}
where $C$ is given by (\ref{C}).
Note that only positive scales $s>0$ are included in the integral (\ref{inv}).  Parseval's
relation (\ref{Parseval}) takes the form
\begin{equation}
\int_{-\infty}^\infty |f(t)|^2\, dt = \frac{2}{C}\int_{0}^\infty \frac{ds}{s^2} \int_{-\infty}^\infty 
dt \,|F(s,t)|^2 .
\label{P}
\end{equation}
Equations (\ref{inv}) and (\ref{P}) are derived in Appendix A.
\medskip

The mother wavelet used in this study is the Morlet wavelet 
\begin{equation}
\psi(t) = \pi^{-1/4} [e^{i\omega_0 t} - e^{-\omega_0^2/2}] e^{-t^2/2},
\label{morlet}
\end{equation}
where, in this study, $\omega_0 = 6$.  Its Fourier transform is
\begin{equation}
\hat\psi(\omega) = 2^{1/2}\pi^{1/4} 
[e^{-(\omega-\omega_0)^2/2}-e^{-\omega_0^2/2}e^{-\omega^2/2}].
\label{psi_hat}
\end{equation}
To leading order, the Morlet wavelet (\ref{morlet}) is a Gaussian envelope modulated by a complex
exponential.  This form is chosen because it is well 
localized in both time and frequency with a small time-frequency uncertainty product.
If small correction terms are introduced to make $\hat\psi(\omega)$ vanish for $\omega<0$,
then the inversion formula (\ref{inv}) may be used.  The magnitude of these correction 
terms are small and can often be neglected in practice. For the Morlet wavelet
(\ref{morlet}), numerical evaluation of the constant $C$ in equation (\ref{C}) yields the value
$C\simeq 1.06$ when $\omega_0 = 6$.
\medskip

To relate the wavelet scale $s$ of the Morlet wavelet to an equivalent Fourier frequency $\omega$, 
recall that the mean frequency $\bar \omega$ of the function $\hat \psi(\omega)$ 
(assumed positive valued) satisfies 
\begin{equation}
\int_{-\infty}^\infty (\omega-\bar \omega) \hat\psi(\omega)  \, d\omega =0.
\end{equation}
Substituting the function (\ref{psi_hat}) into this equation shows that 
$\bar \omega\simeq \omega_0$ is an accurate approximation when $\omega_0\gg 1$.
Because the Fourier transform of the scaled wavelet is proportional to $\hat \psi(s\omega)$,
it follows that the wavelet scale $s$ corresponds to the mean
Fourier frequency  
\begin{equation}
\nu\simeq \frac{\omega_0}{2\pi s}, 
\label{nu}
\end{equation}
where $\nu$ is the frequency in Hz and $\omega=2\pi \nu$ is the frequency in radians per second.
A more accurate approximation is obtained by computing the continuous wavelet
transform of the function $f(t) =\cos(\omega t)$ and then finding the scale $s$ where 
$|F(s,t)|^2$ is a maximum.  To leading order, this yields the refined approximation
derived in appendix B
\begin{equation}
\nu\simeq \frac{\omega_0}{2\pi s}\bigg( 1+\frac{1}{2\omega_0^2}\bigg). 
\label{nu_x}
\end{equation}

\subsection{Practical implementation}

Suppose the function $f(t)$ vanishes outside the interval $0<t<T$.  If $f(t)$ is sampled at the 
discrete times $t_n=n\Delta t$, where $\Delta t=T/N$ and $n=0,1,2,\ldots, N-1$, then the
wavelet transform (\ref{xfm}) can be approximated by the Riemann sum
\begin{equation}
F(s,k\Delta t) \simeq \sum_{n=0}^{N-1} s^{-1/2}\psi^* \bigg[\frac{(n-k)\Delta t}{s}\bigg] 
f(n \Delta t)\, \Delta t.
\label{coeffs}
\end{equation}
This assumes that $\Delta t$ is small enough to resolve the most rapid variations
of the function $f(t)$.  Likewise, the wavelet functions must be sufficiently well 
resolved at all scales $s$ of interest (at least four samples per wave period). Let
$s_m=s_0a^m$, where $s_0$ is the smallest scale, $a>1$, and $m=0, 1, 2,\ldots, M-1$.
Then $\Delta s = s_{m+1}-s_m=(a-1)s_m$ and the inversion formula (\ref{inv}) can be approximated by 
\begin{multline}
f(n\Delta t) \simeq \frac{2}{C}\sum_{m=0}^{M-1} \sum_{k=0}^{N-1} 
\mbox{Re}\, \bigg\{s_m^{-1/2}\psi \bigg[\frac{(n-k)\Delta t}{s_m}\bigg] \\
\times F(s_m,k\Delta t)\bigg\} \frac{(a-1)\Delta t}{s_m}
\end{multline}
For this approximation to be valid the minimum and maximum values of $s$ must include
the smallest and largest timescales of the signal $f(t_n)$ and the grid
size $M$ along the $s$ axis must be chosen large enough to provide adequate frequency
resolution.  To resolve the scaled wavelet function in equation (\ref{coeffs}) requires at least 
four samples per
wavelet period at the minimum scale $s_0$.  This implies $\nu_{\rm max}\le 1/(4\Delta t)$
and, therefore, using equation (\ref{nu}), $s_0 \ge 2\omega_0\Delta t/\pi\simeq 4\Delta t$.  
The maximum scale is typically less than the record length $T$. 
\medskip

From equation (\ref{P}), the total energy of the signal is expressed in terms of the wavelet
coefficients by the Parseval-like relation
\begin{equation}
\sum_{n=0}^{N-1} |f(t_n)|^2\Delta t \simeq \frac{2}{C}\sum_{m=0}^{M-1} \sum_{n=0}^{N-1}  |F(s_m,t_n)|^2 
\frac{(a-1)\Delta t}{s_m}.
\label{Parseval-like}
\end{equation}
The right-hand side of equation (\ref{Parseval-like}) 
describes the distribution of signal energy in the time-frequency plane.
In this study, the minimum scale is $s_0 = 2\omega_0\Delta t/\pi\simeq 4\Delta t$ and the
maximum scale $s_{\rm max}$ is much less than the record length $T$ in which case the
right-hand side of (\ref{Parseval-like}) only includes contributions to the energy
from scales less than or equal to $s_{\rm max}$.  
\medskip

The component of the signal at scale $s_m$ makes a contribution to the total energy 
(\ref{Parseval-like}) given by
\begin{equation}
\frac{2}{C}\sum_{n=0}^{N-1}  |F(s_m,t_n)|^2 \frac{(a-1)\Delta t}{s_m}.
\label{pow}
\end{equation}
The power spectral density $P(\nu)$ is defined such that $P(\nu) d\nu$ is the 
power in the frequency interval from $\nu$ to $\nu +d\nu$.  Using the relation
$d\nu \simeq \omega_0 ds/2\pi s^2$ and the fact that the average power is equal to the
energy divided by the time, equation (\ref{pow})
implies
\begin{equation}
P(\nu_m)=\frac{4\pi \Delta t}{C\omega_0 T}\sum_{n=0}^{N-1}  |F(s_m,t_n)|^2,
\label{psd}
\end{equation}
where the frequency $\nu_m$ is defined by equation (\ref{nu_x}) and $T$ is the record length.  
This expresses the Fourier power spectrum (frequency spectrum) in terms of the wavelet coefficients.
For the vector field $\bm B(t)$, the total power is the sum of the power of
the three orthogonal components so that, in equation (\ref{psd}), 
$|F(s,t)|^2=|F_R(s,t)|^2+|F_T(s,t)|^2+|F_N(s,t)|^2$.
\medskip

In practice, the wavelet coefficients $F(s_m,t_n)$ may be computed using the FFT.
The right-hand side of equation (\ref{coeffs}) consists of a convolution of the sequence
$f_n=f(t_n)$ with the sequence 
\begin{equation}
g_n = \psi^*(-t_n/s).
\label{g_n}
\end{equation}
Note the minus sign in the argument of the function $\psi^*(t)$.  
Recall that for periodic sequences of length $L$, the discrete Fourier transform 
of the convolution  
\begin{equation}
h_n = f_n * g_n=\sum_{m=0}^{L-1} f_m g_{n-m}, \qquad n=0, 1, 2,\ldots, L-1,
\end{equation}
is equal to $H_k=F_kG_k$ \citep{Briggs_Henson:1995}, where 
\begin{equation}
F_k = \sum_{n=0}^{L-1} f_n \exp(-i 2\pi kn/L), \qquad k=0, 1, 2,\ldots, L-1,
\end{equation}
is the discrete Fourier transform of the sequence $f_n$ and the inverse transform is defined by
\begin{equation}
f_n = \frac{1}{L} \sum_{k=0}^{L-1} F_k \exp(+i 2\pi kn/L), \qquad n=0, 1, 2,\ldots, L-1.
\end{equation}
The integer $L> N$ must be chosen to avoid aliasing (wrap around effects) at the largest 
wavelet scale $s$ and is typically a power of 2 to speed up the FFT algorithm.  
The envelope of the scaled Morlet wavelet
$\exp(-t^2/2s^2)$ has an approximate temporal duration of $6s$.
Therefore, the required duration for the zero padding is $3s_{\rm max}$ and $L$ must be
chosen such that $(L-N)\Delta t\gtrsim 3s_{\rm max}$.  Note that the sequence $f_n$ is padded 
with zeros to length $L$, but $g_n$ is not since $g_n$ is defined by equation 
(\ref{g_n}) for $-L/2< n\le L/2$ and for any length $L$.  The wavelet coefficients 
(\ref{coeffs}) at a given scale $s$ are computed as follows.  Compute the FFT of
$f_n$ and $g_n$ at that scale, form the product $F_kG_k$, inverse transform
the sequence $F_kG_k$, and then multiply by $s^{-1/2}\Delta t$.  The first $N$ terms
of the resulting sequence of length $L$ yield the coefficients $F(s,t_n)$
for $n=0, 1, 2,\ldots, N-1$.
\medskip

\begin{figure}
\begin{center}
\includegraphics[width=0.7\columnwidth]{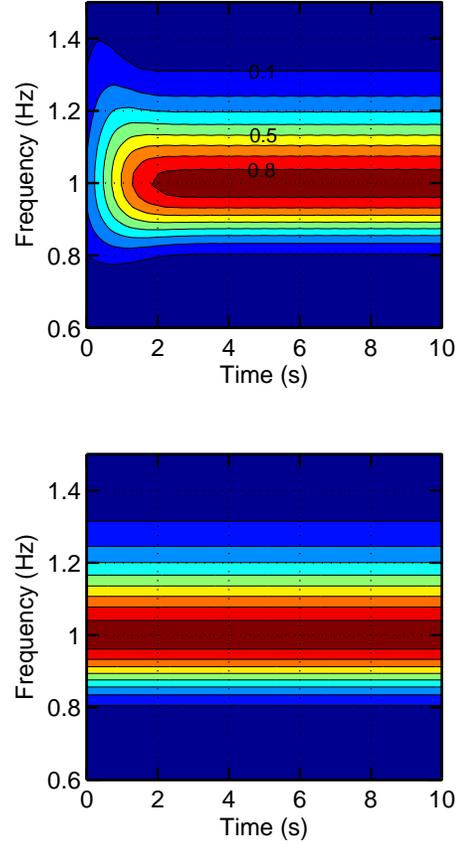}
\caption{\label{fig2}%
The wavelet transform of $f(t)=\cos(\omega t)$ obtained using a Morlet wavelet with $\omega_0=6$.
The numerical results in the upper plot agree with the exact solution in the lower plot.
The transient near $t=0$ (upper plot) is an artifact of the numerical method.
}
\end{center}
\end{figure}

The implementation of the wavelet transform has been tested using synthetic data.  
The wavelet transform of a 1 Hz cosine wave $f(t)=\cos(\omega t)$ is shown in Figure \ref{fig2}.  
The upper panel shows the magnitude of the wavelet transform $|F(s_m,t_n)|^2$ obtained from 
the sequence $f_n=f(n\Delta t)$ using the sampling time $\Delta t=1/200$ sec and a record 
length of 20 sec.
The lower panel shows the exact solution $|F(s,t)|^2$ obtained by evaluating the integral
(\ref{xfm}) in closed form (Appendix B).  Except for the transient near $t=0$, the results for the
wavelet transform agree with the exact solution.  For this particular calculation the minimum 
scale is $s_0 \simeq 8\Delta t$, the maximum scale is $s = 5$ sec, and $M=200$.
The equivalent frequency range extends from $\nu=0.19$ Hz to $\nu=25$ Hz.
The cosine wave is a useful test case and the frequency of the wave is easily changed to
investigate the behavior of the numerical solutions as a function of frequency.
\medskip

\begin{figure}
\begin{center}
\includegraphics[width=0.7\columnwidth]{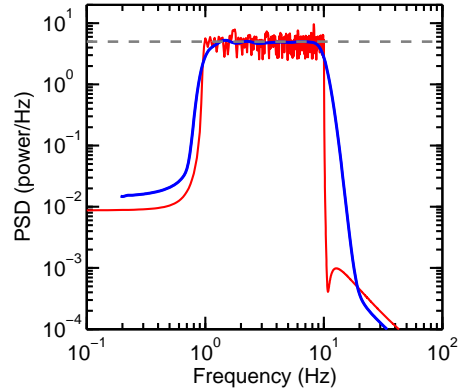}
\caption{\label{fig3}%
The power spectra density (PSD) obtained using the wavelet transform (blue) and the FFT (red) 
for a signal $f(t)$ composed of 91 cosine waves having unit amplitude, random phases, and 
frequencies from 1 Hz to 10 Hz in increments of 0.1 Hz.  The PSD equals 5 Hz$^{-1}$ between
1 Hz and 10 Hz (dashed line).
}
\end{center}
\end{figure}

Many other tests were performed.  One more example shall now be described.  Consider a 20 sec signal 
$f(t)$ composed of a sum of 91 cosine waves having unit amplitude and random phases.  The frequencies 
of the waves extend from 1 Hz to 10 Hz in increments of 0.1 Hz. The power spectrum of $f(t)$
is computed using both the wavelet transform (\ref{psd}) and the FFT which is used to 
compute a frequency spectrum or periodogram (smoothed or unsmoothed).  
The results in Figure \ref{fig3} 
demonstrate the reasonable agreement between these two independent techniques.  For the example
shown in Figure \ref{fig3} , the wavelet transform was computed using the minimum scale 
$s_0 \simeq 4\Delta t$, the maximum scale $s = 5$ sec, and $M=200$.
The FFT spectrum shown in Figure \ref{fig3}  was obtained using a Papoulis smoothing window with a 
bandwidth of 0.1 Hz \citep{Percival_Walden:1993}.  

\section{Analysis of solar wind data}

\indent\indent 
%
The primary goal of this study is to investigate how power is distributed as a function 
of both frequency $\nu$ and the direction $(\theta,\phi)$ of the local mean magnetic field. 
The local mean magnetic field at time $t_n$ and wavelet scale $s$ is obtained by weighting the
time series with a Gaussian curve centered at time $t_n$.  Thus, the mean magnetic field at time 
$t_n$ and wavelet scale $s$ is proportional to
\begin{equation}
\bar{\bm B}_n(s) =\sum_{m=0}^{N-1} \bm B_m \exp\!\left[-\frac{(t_n-t_m)^2}{2\lambda^2s^2}\right]
\label{mmf}
\end{equation}
where $\bm B_n=\bm B(t_n)$.
The width of the scaled Gaussian curve as measured by its full width at $1/e$ of its maximum is 
$2^{3/2}\lambda s\simeq 2.8 \lambda s$.  The dimensionless parameter $\lambda$ determines the timescale 
of the average and in this study $\lambda= 1$.  The local mean magnetic field (\ref{mmf}) 
defines a direction $(\mu,\phi)_n$ at each time step $n$, where $\mu=\cos(\theta)$.  The 
average velocity of the solar wind lies approximately in the heliocentric radial direction
defined as $\theta=0$ or, equivalently, $\mu=1$.
\medskip

Note that the local mean magnetic field in the work of Horbury et al. (2008) does not contain
the factor of 2 which appears in the exponent in equation (\ref{mmf}).  The duration or timescale 
of the average (\ref{mmf}) is therefore slightly larger than that employed by Horbury et al. (2008).
This extraneous factor is not expected to significantly change the results of the analysis.
\medskip

The angular distribution of power may be defined in different ways depending on
the physical quantity of interest.  The power spectral density may be defined 
as a function of frequency and angle such that the integral over all solid angles
and all frequencies is equal to the total power of the signal (energy/time).  However, in the solar wind, 
the direction of the local mean magnetic field is unevenly distributed over the
surface of the unit sphere.  Consequently, for a given
direction $\bm B_0$, the energy at a given frequency
is the product of the {\it average} energy of the fluctuations at that frequency when the
field points in a solid angle $d\Omega$ about the direction $\bm B_0$ multiplied by the number of
times the field $\bm B_0$ points in that direction.  To describe the distribution of
energy of the fluctuations without the weighting caused by the
uneven distribution of directions of the local mean magnetic field, it is necessary
to use the {\it average} energy of the fluctuations at that frequency when the
field points in a given direction (a given angle bin).  This is the approach adopted by
Horbury et al. (2008) and also the approach adopted in this study.  These two distinct approaches
are now described in detail.

\subsection{Angular distribution of power}

The power spectral density (\ref{psd}) may be decomposed into an angular distribution
$P(\nu, \mu,\phi)$ such that the integral over the unit sphere yields the power
spectral density at frequency $\nu$.  In discrete form this is written
\begin{equation}
\sum_{k=0}^{N_\mu-1}\sum_{\ell=0}^{N_\phi-1}P(\nu_m, \mu_k,\phi_\ell)\, \Delta \mu_k\, \Delta \phi_\ell=
\frac{4\pi \Delta t}{C\omega_0 T}\sum_{n=0}^{N-1}  |F(s_m,t_n)|^2,
\label{psd2}
\end{equation}
where $\{\mu_k: k=0, 1, 2,\ldots, N_\mu\}$ is a partition of the interval $(-1,1)$, 
$\{\phi_\ell: \ell=0, 1,\ldots, N_\phi\}$ is a partition of the interval $(0,2\pi)$, 
and the quantity $P(\nu_m, \mu_k,\phi_\ell)\, \Delta \mu_k\, \Delta \phi$
is obtained by including only those terms on the right-hand side for which the direction
$(\mu,\phi)_n$ at time $n$ lies in the interval $\mu_k\le \mu< \mu_{k+1}$ and
$\phi_\ell\le \phi <\phi_{\ell+1}$.  It is also useful to define the angular distribution
$P(\nu, \mu)$ such that
\begin{equation}
\sum_{k=0}^{N_\mu-1}P(\nu_m, \mu_k)\, \Delta \mu_k=
\frac{4\pi \Delta t}{C\omega_0 T}\sum_{n=0}^{N-1}  |F(s_m,t_n)|^2,
\label{psd3}
\end{equation}
where $P(\nu_m, \mu_k)\, \Delta \mu_k$
includes only those terms on the right-hand side for which the direction
$(\mu,\phi)_n$ at time $n$ satisfies $\mu_k\le \mu< \mu_{k+1}$. That is, 
\begin{equation}
P(\nu_m, \mu_k)\, \Delta \mu_k=
\frac{4\pi \Delta t}{C\omega_0 T}\sum_{\genfrac{}{}{0pt}{}{n=0}{\mu_k\le \mu< \mu_{k+1}}}^{N-1}  
|F(s_m,t_n)|^2.
\label{psd_mu}
\end{equation}
The function $P(\nu, \mu)$
describes the distribution of power as a function of the frequency $\nu$ and angle $\theta$
between the mean magnetic field and the average flow direction of the solar wind.
It is the natural way to define the power spectral density per unit frequency and per unit angle $d\mu$.
Consistent with the definition of power spectral density, the total power is equal to the sum 
over all frequencies of $\Delta \nu_m$ times (\ref{psd3}).
Therefore, the contribution to the total power when the 
local mean field lies in a direction between $\mu_k$ and $\mu_{k+1}$ is
the sum over $m$ of $\Delta \nu_m$ times (\ref{psd_mu}), where the mean field depends on the scale $s_m$.
A different definition of the power spectral density is given in the next subsection.

\subsection{Power spectrum as a function of $\theta$}

At a given scale, the number of times the local mean magnetic field $\bm B_0$ points in 
a given direction $\bm r$ within a small solid-angle $d\Omega$ will vary depending on 
the direction of the unit vector $\bm r$.  Thus, the probability distribution of directions of $\bm B_0$
on the surface of the unit sphere is not necessarily uniform.  
By summing the energy of the fluctuations at each time step $t_n$ when $\bm B_0$ points in 
a given direction  $\bm r$, 
the average power of the fluctuations when $\bm B_0$ points in that direction 
is the total energy divided by the length of time $\bm B_0$ points 
in that direction.   This is the power spectral density 
of the fluctuations at a given frequency 
when the local mean magnetic field points in a given direction.
\medskip

Thus, for an angle bin with direction $(\mu,\phi)$ and solid angle $d\Omega \ll 4\pi$, the average power 
per unit frequency when the local mean magnetic field points in that direction $P(\nu, \mu, \phi)$
is defined by
\begin{equation}
P(\nu_m, \mu, \phi)  =\frac{4\pi}{C\omega_0 N_{\rm bin}}
\sum_{\{n:\,(\mu,\phi)_n \in\,\mbox{\scriptsize bin}\}}  |F(s_m,t_n)|^2,
\label{angle_dist}
\end{equation}
where $N_{\rm bin}$ is the number of terms in the sum, that is, the number of times $t_n$ for
which the local mean field lies within the given solid-angle.  The definition (\ref{angle_dist}) 
is believed to be equivalent to that employed by Horbury et al. (2008) and it is also the definition
used in the present study.  
\medskip

Note that the usual Fourier analysis of solar wind time series ignores the fact that the 
local mean field spends different lengths of time pointing in different directions. This uneven 
distribution of directions of $\bm B_0$ on the surface of the unit sphere causes an uneven 
weighting of power in a standard Fourier spectral analysis.  The spectral analysis 
based on equation (\ref{angle_dist}) removes this uneven weighting by dividing by the number of times
the local mean field points in a given direction.  
This is made possible by the fact that the wavelet transform is localized in both time 
and frequency and, in this respect, the wavelet transform is superior to the Fourier transform.
\medskip

The solar wind data listed in Table 1 was analyzed by means of the wavelet transform 
(\ref{coeffs}) and the spectral decomposition (\ref{angle_dist}).  The
analysis shows that for frequencies throughout the inertial range
and the dissipation range, frequencies from $5\times 10^{-3}$ Hz to 2 Hz in the spacecraft frame, 
the power spectral density
in equation (\ref{angle_dist}) is roughly independent of the azimuthal angle $\phi$ as is
expected for turbulence that is cylindrically symmetric about the local mean magnetic field.
Therefore, instead of equation (\ref{angle_dist}), it is more expedient to
use the power spectral density as a function of angle defined by
\begin{equation}
P(\nu_m, \mu_k)=
\frac{4\pi}{C\omega_0 N_{mk}}\sum_{\genfrac{}{}{0pt}{}{n=0}{\mu_k\le \mu< \mu_{k+1}}}^{N-1}  
|F(s_m,t_n)|^2,
\label{psd_theta}
\end{equation}
where the sum on the right-hand side only includes those terms for which the direction
$(\mu,\phi)_n$ at time $n$ satisfies $\mu_k\le \mu< \mu_{k+1}$ and  $N_{mk}$ is the 
number of all such terms.  The function $P(\nu, \mu)$
describes the power spectral density as a function of the angle $\theta$
between the local mean magnetic field and the (radial) mean flow direction.
\medskip

In this study, angle bins are defined by the partition $\theta_k=k\pi/30$,
$k=0, 1, 2,\ldots, 30$, $\mu_k=\cos(\theta_k)$.  The width of each angle bin is 6 degrees
which is small enough to resolve the interesting behavior near the endpoints 
$\theta=0$ and $\theta=\pi$,
but large enough to usually provide a reasonable statistical sample at all angles of interest.
\begin{figure}
\begin{center}
\includegraphics[width=0.8\columnwidth]{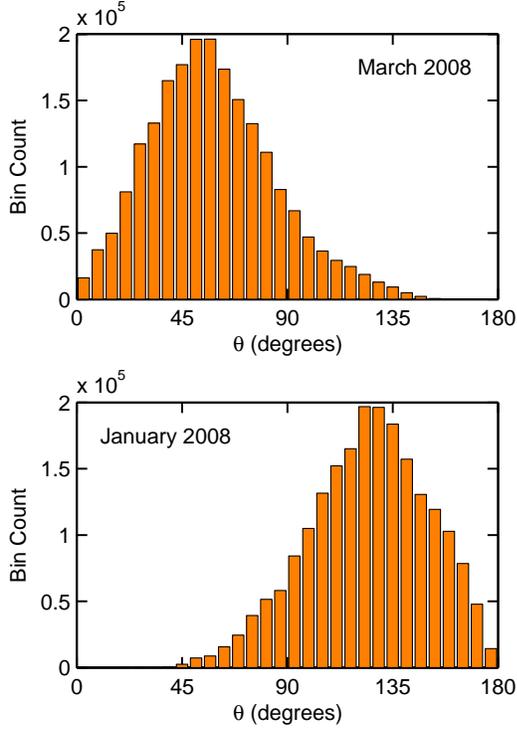}
\caption{\label{bin_counts}%
The number of bin counts for each angle bin at the frequency 0.716 Hz using
the data from March 2008 and January 2008  in Table 1.  The direction $\theta=0$ is
radially outward from the sun.  
Note that these two data sets are contained in opposite magnetic sectors (inward and outward).
}
\end{center}
\end{figure}
Figure \ref{bin_counts} is a typical example of the distribution of bin counts seen for the data in 
this study.  The orientation of the interplanetary magnetic field, outward or inward, is
also apparent from Figure \ref{bin_counts}.  The maximum bin count coincides with the direction
of the Parker spiral in each case.  The distributions of bin counts was qualitatively and 
quantitatively similar at all frequencies from $5\times 10^{-3}$ Hz to 2 Hz.  The 
number of bin counts in the parallel direction was typically 4\%--9\% of the 
maximum number of bin counts taken over all bins.

\section{Results}

\begin{figure}
\begin{center}
\includegraphics[width=0.8\columnwidth]{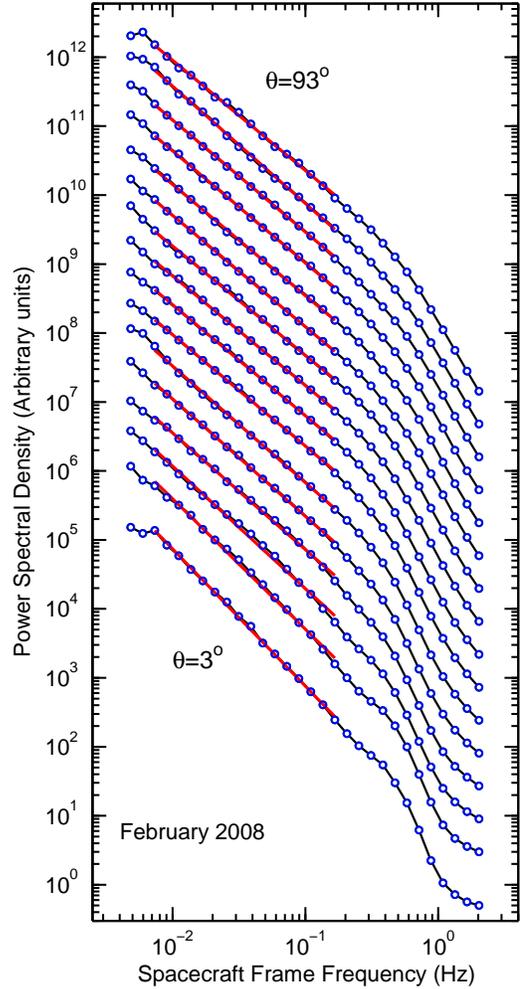}
\caption{\label{spectra}%
The power spectral density versus frequency for angle bins centered at $\theta=3$ (bottom), 
9, 15, 21,$\ldots,$ 93 degrees (top) computed using the February 2008 data 
in Table 1 by means of equation (\ref{psd_theta}).
}
\end{center}
\end{figure}

\indent\indent 
The power spectra $P(\nu, \mu)$ computed by means of equation (\ref{psd_theta})
using the February 2008 data in Table 1 are shown in Figure \ref{spectra}.
The different curves in Figure \ref{spectra} correspond to different
angle bins and have been offset vertically for easier viewing.  
The minimum and maximum wavelet scales are approximately 0.48 s and 200 s, respectively.
The number of different scales is $M=30$.
The corresponding frequency range extends from approximately $5\times 10^{-3}$ Hz to 2 Hz
in the spacecraft frame.   The transition from the
low-frequency inertial range to the high-frequency dissipation range is indicated by the change in
spectral slope around 0.4 Hz, a typical break-frequency near the orbit of the earth at 1 AU.
\medskip

Restricting attention to the range of frequencies called the inertial range, 
$\nu\lesssim 0.2$ Hz, linear least squares fits are performed in log-log space to find the best fit
power-law exponents.  For a given data record, the frequency range used to obtain the fits is 
the same for all angles. But, for each data record in Table 1 a different
frequency range is used to fit the data.  This was necessary
because of occasional outliers in the spectra, usually at the lowest frequencies, 
which can significantly effect the power law fits.  Such outliers are partly attributable 
to the smaller bin count at extreme angles and low frequencies.
The frequency ranges used to fit the data are listed in Table 1.  An example of the power-law fits
are shown by the red line segments in Figure \ref{spectra} which have been drawn so that they cover the precise
frequency interval used to determine the fit.
\medskip

\begin{figure*}
\begin{center}
\includegraphics[width=4.5in]{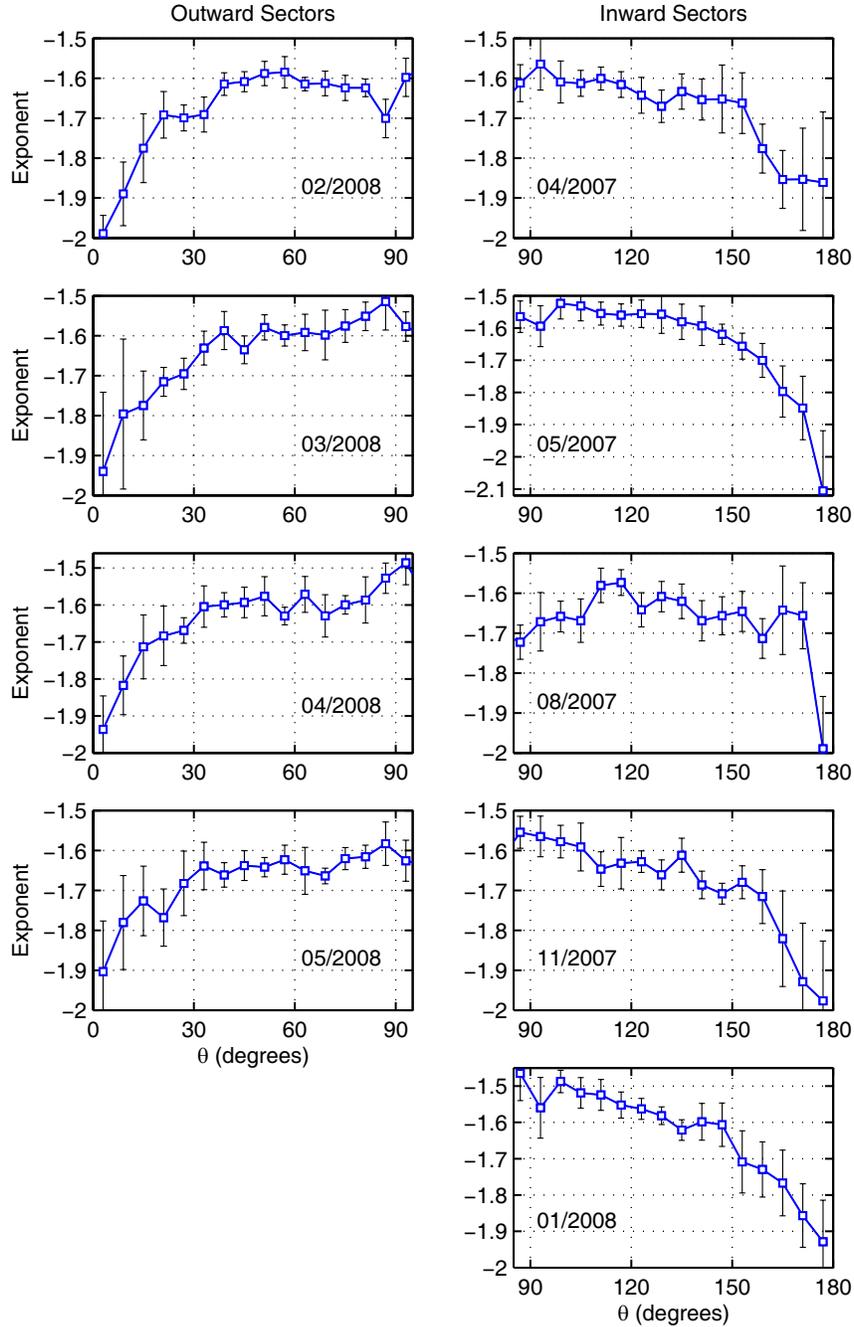}
\caption{\label{exponents}%
Power-law exponents as a function of the angle $\theta$ between the direction of the
local mean magnetic field and the heliocentric radial direction (approximately the
mean flow direction).  
}
\end{center}
\end{figure*}

The dependence of the inertial range power-law exponents on the angle $\theta$ are
shown in Figure \ref{exponents}.  
For outward magnetic sectors, the power-law exponent changes from 
roughly 2 to 1.6 as the angle $\theta$ increases from 0 to 90 degrees.  The results are 
qualitatively similar in every case.  For inward magnetic sectors, the power-law exponent 
changes from roughly 2 to 1.6 as the angle $\theta$ decreases from 180 to 90 degrees.  
Once again, the results are qualitatively similar in every case.
The error bars for the power law exponents in 
Figure \ref{exponents} are 99\% confidence intervals based on linear regression
analysis of the data on a log-log plot, they are not standard deviations.  
\medskip

Near $\theta=0$ in outward sectors and 180 degrees in inward sectors the power-law exponents 
are usually more uncertain as a consequence of the smaller number of 
statistical samples at these extreme angles.  
Nevertheless, the data indicate that
the power law exponent when the mean field is parallel to the mean flow is 
approximately $2\pm 0.1$.    The value of 
the power-law exponent when the mean field is perpendicular to the mean flow is 
approximately $1.6\pm 0.1$.  In some cases the perpendicular power-law exponent is closer
to 3/2 than 5/3, but the observations show significant variations from one angle bin to the next
and from one data record to another
so it is difficult to determine this value more precisely.  Statistical studies may help answer
this important question.
\medskip

\begin{figure}
\begin{center}
\includegraphics[width=0.8\columnwidth]{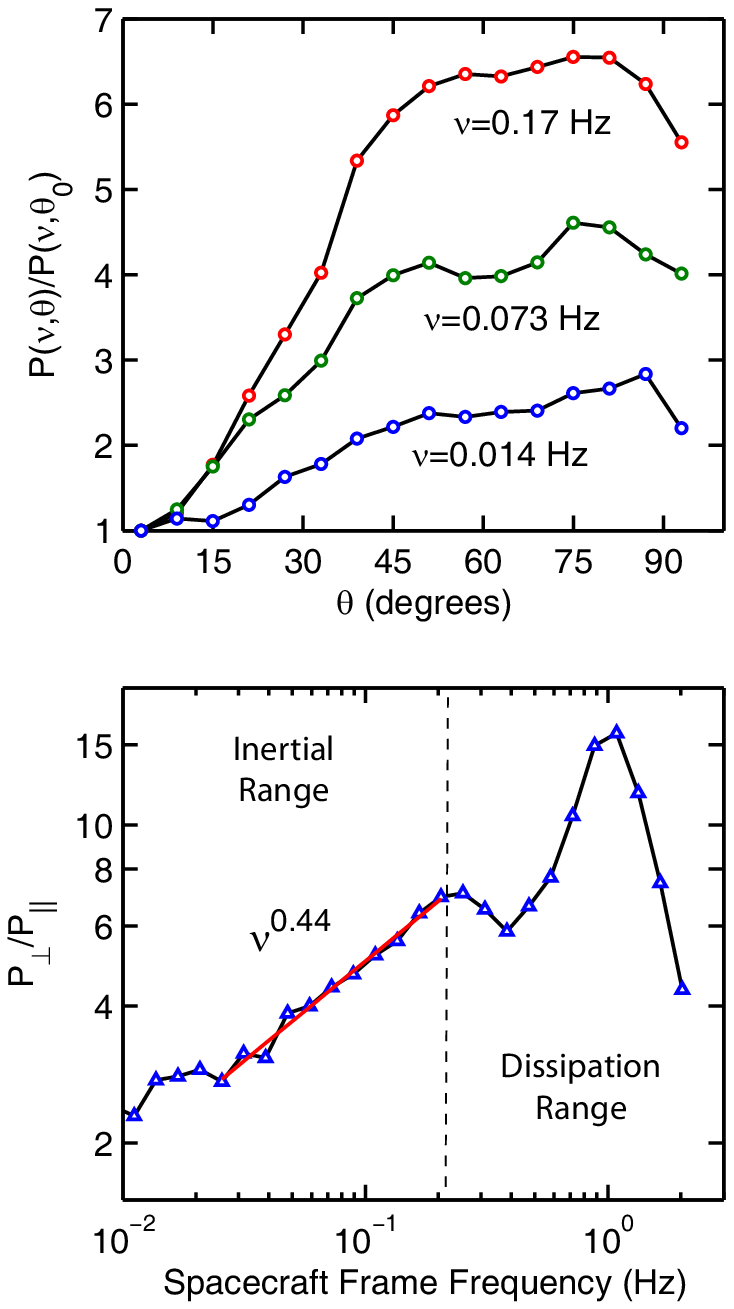}
\caption{\label{p_vs_angle}%
The power spectral density $P(\nu,\theta)$ normalized to the value in the
parallel direction $\theta_0=3$ degrees and plotted as a function of the angle $\theta$
between the measurement direction and the direction of the local mean magnetic field
for three different inertial range frequencies (upper plot); February 2008 data.  
The anisotropy described by the ratio of the perpendicular to the parallel power, $P_\perp/P_\parallel$,
increases as the frequency increases through the inertial range and a well defined peak is seen in 
the dissipation range at 1 Hz (lower plot). The red line fit is proportional to
$\nu^{0.44}$.
}
\end{center}
\end{figure}

The power spectral density $P(\nu,\theta)$ as a function of angle $\theta$ is shown in the upper
plot in Figure \ref{p_vs_angle} 
which shows that the power is peaked near the perpendicular 
direction and that the anisotropy of the fluctuations described by the ratio
of the perpendicular to the parallel power $P_\perp/P_\parallel$ increases as the frequency increases
through the inertial range.  The inertial range observations in Figure \ref{p_vs_angle} 
are consistent with direct numerical
simulations of incompressible MHD turbulence which show that in $\bm k$-space the energy 
cascade is directed primarily perpendicular to
the mean magnetic field and that the power anisotropy increases as the wavenumber increases
\citep{Shebalin:1983, Oughton:1994, Matthaeus_Ghosh:1996}.  This behavior is also incorporated
into phenomenological theories of strong incompressible MHD turbulence that
take wavevector anisotropy into account in a fundamental way
\citep{Goldreich_Sridhar:1995, Goldreich_Sridhar:1997,
Boldyrev:2005, Boldyrev:2006, Lithwick_Goldreich:2007, Beresnyak_Lazarian:2008, Chandran:2008,
Podesta_Bhattacharjee:2009}.
\medskip

In the lower plot in Figure \ref{p_vs_angle}, the quantity $P_\perp$ is the average power in the
two angle bins just below 90 degrees, bins 14 and 15, and $P_\parallel$ is the value in bin 1.
The ratio $P_\perp/P_\parallel$ found in the present study  
is quantitatively similar to the results of \citet{Bieber:1996} 
who found the ratio 1.4 (mean value) at 0.01 Hz
and to the results of Horbury et al. (2008) who found a ratio of between 4 and 5 at 0.061 Hz
at high heliographic latitudes near 1.4 AU.   Smith et al. (2006) decomposed the magnetic field
vector into orthogonal components perpendicular and parallel to the mean field and found
that the average ratio of the power spectral density of the sum of the two perpendicular components 
to the parallel component, averaged from $8\times 10^{-3}$ Hz to 0.1 Hz, takes typical values around
10.  It is not suprising that this {\it variance anisotropy} measured by \citet{Smith:2006} 
is quantitatively similar to the ratio $P_\perp/P_\parallel$ studied here, although the two 
physical quantities are different.  Measurements similar to those of \citet{Smith:2006}
were also reported by \citet{Leamon:1999b}
who found a mean ratio around 10 at the high frequency end of the inertial range.
\medskip

An interesting new result is the observation that
in the turbulent dissipation range the power ratio $P_\perp/P_\parallel$ often exhibits a
double peak structure as shown in the lower plot in Figure \ref{p_vs_angle}.
What is usually called the dissipation range begins at the spectral break 
(change in slope) in Figure \ref{spectra}, the point where the spectra in Figure \ref{spectra}
deviate from their linear fits, which occurs around 0.25 Hz for the data 
from February 2008.  For the same data,
the lower plot in Figure \ref{p_vs_angle} shows that the ratio $P_\perp/P_\parallel$ decreases,
attains a local minimum at around 0.4 Hz, and then increases again.  What is most
interesting is that it exhibits a prominent  
peak of large magnitude near 1 Hz before decreasing significantly beyond 1 Hz.  
This feature has never been observed before.  A prominent peak near 1 Hz is seen
in all the intervals listed in Table 1.
\medskip


According to one school of thought, the turbulent energy cascade of MHD-scale fluctuations 
makes a transition from an Alfv\'en wave cascade to a Kinetic Alfv\'en Wave (KAW) cascade 
at the perpendicular 
wavenumber $k_\perp \rho_i\simeq 1$ \citep{Quataert:1998,Quataert:1999,
Leamon:1999c,Cranmer:2003,Howes:2008a,Howes:2008b,Schekochihin:2008,Schekochihin:2009}.  
Observational evidence for such a transition in the solar wind is found in 
the measurements of \citet{Bale:2005} which show the phase speed of the waves in this regime 
are consistent with the dispersion relation for KAWs.
For the February 2008 data in the present study, the wavenumber such that
$k_\perp \rho_i\simeq 1$ occurs at the observed spacecraft frequency 
$\nu_{\rm sc}=V_{\rm sw}/\lambda_\perp\simeq 0.5$ Hz.
Thus, the onset of the KAW cascade occurs near the spectral break ($\sim 0.3$ Hz) and also near
the location of the local minimum in the lower plot in Figure \ref{p_vs_angle}.  The observations 
in the lower plot in Figure \ref{p_vs_angle} show that the ratio $P_\perp/P_\parallel$
increases as the frequency increases from 0.5 Hz, reaches a peak around 1 Hz, 
and then decreases rapidly near 2 Hz.  The observed peak with the property 
$P_\perp/P_\parallel\gg 1$ demonstrates the perpendicular
nature of the cascade in the dissipation range which is consistent with the
existence of a kinetic Alfv\'en wave cascade in this regime.
\medskip

The rapid decrease in the power ratio $P_\perp/P_\parallel$ beyond 1 Hz could be due to the
dissipation of kinetic Alfv\'en waves (KAWs).  The linear dispersion relation for 
electromagnetic waves in a collisionless electron-proton plasma at thermal equilibrium 
contains two branches which correspond, in the limit $k_\perp \rightarrow 0$, to the 
left circularly polarized ion-cyclotron (Alfv\'en-cyclotron) wave and the right circularly 
polarized electron-cyclotron (magnetosonic-whistler) wave.  Suppose that the 
magnetosonic-whistler can be neglected in the first order of approximation.
For the relevant solar wind parameters $\beta_{\perp p}=2$, $v_{th,p}=75$ km/s, and 
$k_\perp/k_\parallel=10$,  
the hot plasma dispersion relation shows that the damping of KAWs becomes strong, 
$\gamma T\simeq -1$, when $k_\perp \rho_i\simeq 4$.  This corresponds to an observed 
frequency of 2 Hz where, assuming the wave amplitudes are sufficiently small, the 
kinetic Alfv\'en waves are extinguished.  
Strong damping of KAWs near 2 Hz may cause the attenuation of the peak  
in the lower plot in Figure \ref{p_vs_angle}.  However, another contributing factor may be a 
population of parallel propagating waves as explained later in this section.  
\medskip

The measured plasma parameters used to make the above estimates for the February 2008 data 
in Table 1 consist of the average solar wind speed $V_{\rm sw}=625$ km/s, proton density 
$n_p=3$ cm$^{-3}$, proton thermal speed  $v_{th,p}=75$ km/s, magnetic field magnitude 
$B=4.2$ nT, and the Alfven speed $V_A=49$ km/s.  Measurements of the electron core 
temperatures on the Stereo spacecraft are unavailable due
to instrument problems (P. Schroeder, private communication 2008).  The ratio $T_p/T_e=2$
is a typical value in high-speed wind \citep{Marsch:1991b, Schwenn:2006,Marsch:2006},
however, here it is assumed for simplicity that $T_p/T_e=1$.
The ratio $k_\perp/k_\parallel=10$ in the last paragraph comes from the assumption
that MHD-scale turbulence in the solar wind obeys the critical balance hypothesis so that
$k_\parallel/k_\perp \sim \delta v_k/v_A$;  solar wind measurements indicate that at the
smallest inertial range scales  $\delta v_k/v_A\sim 1/10$.  The exact dispersion 
relation given by equations (10-57) and (10-66) in \citet{Stix:1992} were used to
compute the damping rates and Gautschi's algorithm was used to
compute the plasma dispersion function \citep{Gautschi:1970}.  The root of the
dispersion relation where the damping becomes strong, $\gamma T\simeq -1$, is
given by $(\omega/\Omega_p)/(k_\parallel \rho_i)=1.5602-0.2375i$, where 
$k_\parallel \rho_i=0.4$, $k_\perp \rho_i=4$, and the proton cyclotron frequency 
is $\Omega_p/(2\pi)=0.064$ Hz.
It is known that damping coefficients can be sensitive
to small changes in the distribution functions which are usually far from 
thermal equilibrium in the solar wind.  Therefore, the above calculation of damping 
coefficients of KAWs in the solar wind is subject to possibly large errors.
\medskip

Another noteworthy feature in the lower plot in Figure \ref{p_vs_angle} is the 
decrease of the ratio $P_\perp/P_\parallel$ around the spectral break and the local minimum 
near 0.4 Hz.  This arises from an increase in the power $P_\parallel$ measured parallel to 
$\bm B_0$ relative to the power $P_\perp$ measured perpendicular to 
$\bm B_0$ and can also be seen in Figure \ref{spectra} where it appears as a slight
enhancement in the power spectrum near the spectral break in the bottom two curves.  
The probable cause is an enhanced population of nearly parallel propagating waves with
wavenumber $k_\parallel \rho_i \sim 1$.  Because the enhancement in the power spectra is only
seen at small angles in Figure \ref{spectra}, $\theta \ll \pi/2$,  it is suspected that
the waves are parallel propagating.  More extensive data analysis is necessary to
confirm this.  At the present time, the source of the waves is unknown.
It is not known if they are part of the turbulence, somehow generated by the turbulence,
or whether they are independent of the turbulence.  The waves could be generated in-situ by 
a kinetic instability.  Coherent, parallel propagating waves around 0.5 Hz were 
frequently detected in the solar wind by \citet{Behannon:1976} who tentatively identified both 
ion-cyclotron and electron cyclotron waves in different data sets; similar
observations have been made by \citet{Jian:2008}.    
Assuming the instability only exists for a narrow range of $k_\parallel$,
then as $k_\parallel$ increases beyond this range $P_\parallel$ would be expected to 
decrease rapidly and this may explain why the ratio $P_\perp/P_\parallel$ in the 
lower plot in Figure \ref{p_vs_angle} ramps up rapidly 
to form the peak near 1 Hz.  These possibilities require further investigation.
\medskip

Remarkably, a similar enhancement of proton
density fluctuations in the solar wind near $k_\parallel \rho_i \sim 1$ has been reported by 
\citet{Neugebauer:1975, Neugebauer:1976}.  \citet{Hollweg:1999} has suggested these enhanced 
density fluctuations could be caused by the compressibility of KAWs in the case 
$k_\perp\gg k_\parallel$.  However, if the enhancement in the present study is 
caused by {\it parallel} propagating waves, then it cannot be explained by Hollweg's 
mechanism since, for parallel propagating waves, $k_\perp\ll k_\parallel$.
\citet{Neugebauer:1978} have suggested that the enhancement could be caused by 
right-hand ion cyclotron waves driven unstable by a proton thermal anisotropy
with $T_{\parallel p}>T_{\perp p}$.  While this could also explain 
the wavelet observations in the present study,
other kinetic instabilities are also possible.
The right-hand electromagnetic (whistler) wave instability driven by a proton beam
\citep{Montgomery:1976} was dismissed by Neugebauer et al. (1978) because they believed 
in this case the unstable waves exist over a frequency range that is wider than 
what is seen in observations.  The crucial quantity, however, is not the frequency in the
plasma frame but the {\it wavelength} because the frequency observed in the spacecraft frame
is determined by the rapid advection of the waves past the spacecraft (by the solar wind)
and, therefore, is determined by the wavelength. Thus, 
by Taylor's hypothesis, $\nu_{\rm sc}=V_{\rm sw}/\lambda$.
Figure 4a in \citet{Montgomery:1976} shows that the wavenumber range of the
unstable waves could explain the observed density and/or magnetic field 
enhancements near $k_\parallel \rho_i \sim 1$. 
\medskip

The rapid attenuation of the peak around 2 Hz is another characteristic feature of the 
dissipation range data in the lower plot in Figure \ref{p_vs_angle} that could be 
caused by an enhanced population of predominantly parallel propagating waves. 
Evidence comes from the power spectra in Figure \ref{spectra} which show a ``flattening'' 
of the spectra near 2 Hz or, equivalently, $k_\parallel \rho_i \sim 4$.  This flattening is
more pronounced at small angles but is present in angle bins 1 through 8 indicating that 
if such waves are present, then they may propagate at angles as large as $\theta=45$ 
degrees or so, even though the wave power is greatest for propagation near $\theta=0$.  
It should be noted that the flattening of the power spectra cannot be caused by aliasing
which would affect the spectra at all angles in the same way.  Moreover,
the underlying measurements were obtained with a cadence of 32 vectors per second.
Further analysis of the data is necessary to investigate whether parallel propagating 
waves are present around 2 Hz and, if present, to characterize their properties 
which may be similar to those identified by Behannon (1976).
\medskip

Moving now to the inertial range, an interesting observation in the lower plot in Figure 
\ref{p_vs_angle} is the red trend line with slope 0.44 which is the best least-squares fit 
to the inertial range data between 0.025 Hz and 0.2 Hz.  As discussed in the conclusions, the
slope 1/2 is predicted by the phenomenological theory of Boldyrev (2006) and
the slope 1/3 is predicted by the theory of Goldreich \& Sridhar (1995).  
The best least-squares fit over the range from 
0.01 Hz to 0.2 Hz yields the
exponent 0.36 (not shown), however, it is not as good a fit as the exponent 0.44
over the range 0.025 Hz to 0.2 Hz.  
These considerations show that the scaling law is difficult to determine from the data.
This difficulty is caused in part by the
uncertainties in the ratio $P_\perp/P_\parallel$ which increase as the frequency decreases. 
These uncertainties can be reduced by using larger record lengths, but the
record length is limited by the duration of the high speed stream. 
Further study is needed of this important scaling law.

%

\section{Comparison to Ulysses results}

It is of interest to compare the results of the present study with those obtained by
Horbury et al. (2008).  Unfortunately, the numerical results of Horbury et al. (2008) were 
not made availible to us.  Therefore, the analysis of
Horbury et al. (2008) was repeated using the similar analysis technique developed in
the present work. 
\medskip

The data consists of measurements of the solar wind magnetic field by the vector helium
magnetometer on the {\it Ulysses} spacecraft \citep{Balogh:1992}. The data spans a 30 day 
interval in 1995, day 100 through day 129, the same interval analyzed by Horbury et al. (2008).  
The data has a nominal 1 second cadence, however, some 1/2 s data and data gaps of 
various sizes are also present.  The 1/2 s data is downsampled to 1 s and all data 
gaps are filled using linear interpolation
to give a continuous time series with a 1 second cadence.  
This is necessary to compute convolutions using FFT techniques as described in section 4.  
The duration of the data gaps is negligible compared to the record length and, therefore, 
the linearly interpolated data makes a negligible contribution to the total power.  
The mean value of $\bm B=(B_R, B_T, B_N)$ taken over the
entire record is $(1.65,-0.74,-0.03)$ nT.  The mean value of $B=|\bm B|$ is 3.07 nT and
the r.m.s.\ value of $\bm B$ is $\delta B =2.53$ nT.  Thus, for the 
{\it Ulysses} data $\delta B/B\sim 1$.
For each of the time intervals listed in Table 1, $\delta B/B\simeq 0.8$.
\medskip

\begin{figure}
\begin{center}
\includegraphics[width=0.8\columnwidth]{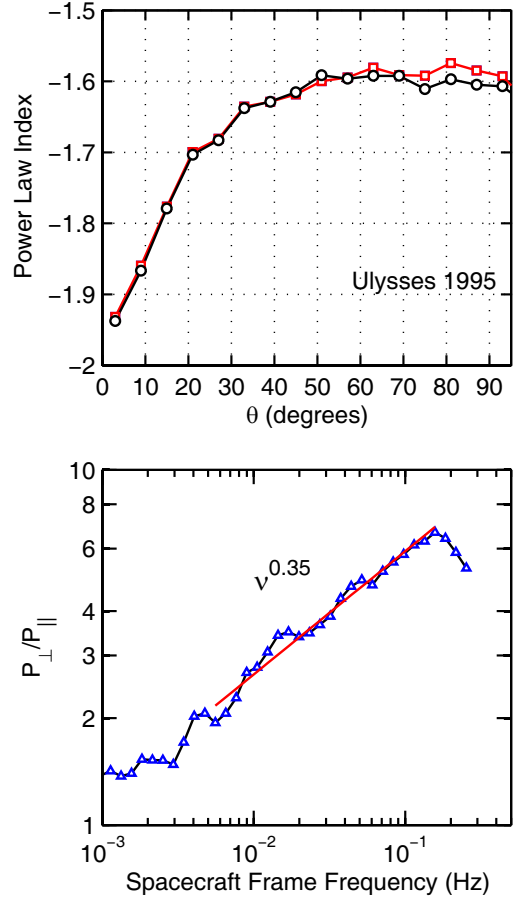}
\caption{\label{Ulysses}%
The power law exponents as a function of the angle $\theta$ between the local mean
magnetic field and the mean flow direction (radial direction) for the 1995 Ulysses data,
DOY 100--130 (upper plot).  The black curve is obtained by averaging the power in each direction 
$(\mu,\phi)$ separately and then performing a $\phi$-average (upper plot).  The red curve is obtained by 
averaging with respect to $\phi$ only which yields better statistics (upper plot).
The ratio of the perpendicular to the parallel power, $P_\perp/P_\parallel$,
increases approximately like  $\nu^{1/3}$ as predicted by the Goldreich \& Sridhar
theory (lower plot). The linear least-squares fit yields the slope $0.35\pm 0.04$.
}
\end{center}
\end{figure}

The wavelet analysis was performed in the same manner decribed in sections 4 and 5.  
Horbury et al. (2008) average the energy in each solid angle bin and then perform an
azimuthal average with respect to $\phi$.  In the present study, intermediate averages over each
solid angle bin were omitted and, in accordance with equation (\ref{psd_theta}), only 
averages with respect to $\phi$ are performed.  This has the advantage of yielding
more robust statistics for angles in the parallel and perpendicular directions
where the bin counts are relatively small.  The results of the Ulysses analysis
are shown in Figure \ref{Ulysses}.  
\medskip

Overall, the Ulysses results for the power-law indices in Figure \ref{Ulysses} are qualitatively 
and quantitively similar to the results for high-speed streams in the ecliptic plane shown in 
Figure \ref{exponents}.
The power-law exponent closest to $\theta=0$ in Figure \ref{Ulysses} is not as close to 2 as in the
results reported by Horbury et al. (2008), possibly because the bin size used here is 
larger.  The power-law exponents approach values near 1.6 in the perpendicular direction,
notably different than the value 5/3 reported by Horbury et al. (2008).  When 
intermediate averages over each
solid angle bin are omitted and averages are performed only with respect to $\phi$, as in 
equation (\ref{psd_theta}), the results take the form of the red curve in the upper plot in 
Figure \ref{Ulysses}.  Note that the red curve gives a more monotonic functional form 
with a perpendicular power law exponent closer to 1.58 than 1.67.  The precise value
of the parallel and perpendicular power-law exponents are important for comparisons with
turbulence theories.  
\medskip

The ratio of power in the perpendicular direction to the power in the parallel direction,
$P_\perp/P_\parallel$, is shown in the lower plot in Figure \ref{Ulysses}.  Here,
$P_\parallel$ is the power in angle bin 1, $P_\perp$ is the average power in 
bins 14 and 15, and the power is defined by equation (\ref{psd_theta}).
As described in more detail in the conclusions, the Goldreich \& Sridhar (1995) theory predicts
$P_\perp/P_\parallel \propto \nu^{1/3}$ and the Boldyrev (2006) theory predicts
$P_\perp/P_\parallel \propto \nu^{1/2}$.  The data in the lower plot in Figure \ref{Ulysses}
yield the power law fit $\nu^{0.35\pm 0.04}$.  This result is consistent with 
the Goldreich \& Sridhar (1995) theory but inconsistent with the Boldyrev (2006) theory.  
These results are new and were not analyzed this way by Horbury et al. (2008).
\medskip


\section{Comparison to power spectrum using all data}

\begin{figure}
\begin{center}
\includegraphics[width=0.9\columnwidth]{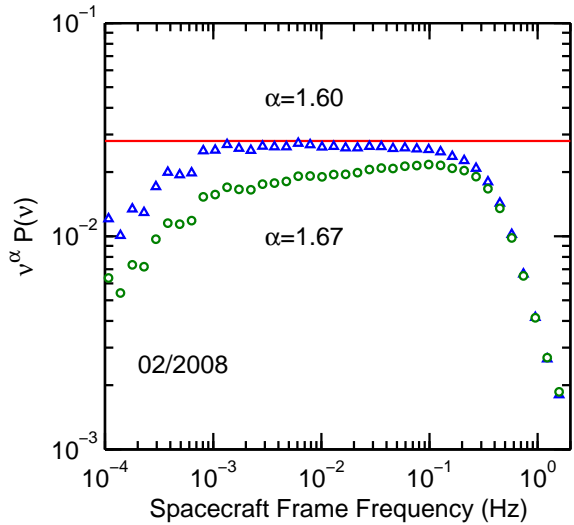}
\caption{\label{spectrum_all_data}%
The compensated power spectrum $\nu^\alpha P(\nu)$, trace spectrum, obtained using data 
for February 2008 in Table 1.  The best fit power-law exponent over the frequency interval
from $10^{-3}$ Hz to $10^{-1}$ Hz is $\alpha=1.60$ (triangles).  The exponent $\alpha=5/3$ is
not a good fit (circles).  The horizontal line is drawn for comparison.
}
\end{center}
\end{figure}

It is of interest to compare the results for the power law exponents in section 5 
to the power law exponents obtained using all data.  The power spectral density for each 
record (trace power) is computed using equation (\ref{psd}) which yields the same result as 
the power spectrum computed using standard FFT techniques \citep{Percival_Walden:1993}.
The spectral exponent is obtained by a linear least squares fit over the interval
from $10^{-3}$ Hz to $10^{-1}$ Hz.  An example of the compensated spectrum for the
February 2008 data in Table 1 is shown in Figure \ref{spectrum_all_data}.  
The power law fits for the other inervals are qualitatively similar to that shown in
Figure \ref{spectrum_all_data}.  
A summary of the power law exponents for each interval are listed in Table 1.
\medskip

The power law exponent for each interval (all data) is in approximate agreement with the power-law 
exponents observed over the range of angles
between $\theta=\pi/2$ and the direction of the Parker spiral.  This is to be
expected because observations in Figure \ref{exponents} shown that the power law exponent is 
approximately constant 
between $\theta=\pi/2$ and the direction of the Parker spiral and the amplitude of the
power spectrum is a maximum roughly when $\theta=\pi/2$.  Furthermore, the fact that the 
direction of the local
mean magnetic field is most often found near the direction of the Parker spiral means that
the Fourier spectrum for all data is more heavily weighted by those fluctuations occuring when $\bm B_0$
lies near the Parker spiral direction.  Unfortunately, the large uncertainties for the
exponents in Figure \ref{exponents} preclude more precise comparisons of the power law exponents 
as a function of angle $\theta$ and the power law exponents obtained using all data.

\section{Discussion and conclusions}

\subsection{Summary}

Wavelet analysis of solar wind data obtained at high heliographic latitudes by the Ulysses 
spacecraft led Horbury et al. (2008) to conclude that in the inertial range the power-law exponent of 
solar wind magnetic field fluctuations changes from approximately 5/3 to approximately 2
as the angle $\theta$ between the local mean magnetic field and the mean flow direction
decreases from $\pi/2$ to 0.   The similarity between high-speed solar wind at high latitudes
and high-speed streams in the ecliptic plane around solar minimum suggests that a similar
effect should exist for power-law exponents in high-speed streams.  The results of this study 
confirm this expectation.  
\medskip

Each of the nine high-speed streams analyzed in this study showed a decrease in
the power-law exponent from $2\pm 0.1$ to $1.6\pm 0.1$ as the angle $\theta$ increases
from 0 to $\pi/2$ for streams embedded in outward-directed magnetic sectors (large scale 
$\bm B_0$ pointing away from the sun) and as the angle $\theta$ decreases
from $\pi$ to $\pi/2$  for streams embedded in inward-directed magnetic sectors 
(large scale $\bm B_0$ pointing toward the sun).  
\medskip

In the inertial range, observations show that the power in the perpendicular direction 
$\theta=\pi/2$ is greater than the power in the parallel direction $\theta=0$ or $\theta=\pi$
by factors ranging from approximately 2 at 0.01 Hz, near the middle of the inertial range 
to a factor of roughly 10 at 0.2 Hz, the upper end of the MHD-scale inertial range.  
Thus, it may seem that the power spectrum of the entire data record should have 
approximately the same power-law exponent as the spectrum in the perpendicular direction 
$\theta=\pi/2$.  If true, then this provides a way to obtain
more accurate estimates of the power-law exponent when $\theta=\pi/2$.  However, as
discussed in section 4.2, because
the usual Fourier frequency spectrum is weighted by an uneven distribution of directions of the
local mean magnetic field which is peaked in the direction of the Parker spiral, the power 
spectrum of the entire data record will likely have a power-law exponent that lies between 
the values observed in the directions $\theta=\pi/2$ and the direction of the Parker spiral,
and probably closer to that in the direction of the Parker spiral because of its 
heavier statistical weight.
\medskip

Interestingly, this effect could cause a spectrum with a perpendicular power-law exponent
near 3/2, as measured using the wavelet technique at $\theta=\pi/2$, to produce a power-law 
exponent that is close to 1.6 or even 5/3 as measured by Fourier spectral analysis of the entire record.
In fact, the results in Table 1 and in Figure \ref{spectrum_all_data} show that the power law 
exponent for the total magnetic energy is closer to 1.60 than 1.67 in all the inervals studied here.
Therefore, it is possible that Fourier analysis techniques and Fourier frequency spectra which have been
a standard tool in solar wind research for five decades may be giving misleading results
for the power law index of turbulent MHD-scale fluctuations. 

\subsection{Comparison to previous work}

It is of special interest to comment on the large statistical study by 
\citet{Tessein:2009} who examined both high- and low-speed wind in the ecliptic plane
near 1 AU and found no statistically 
significant dependence of the spectral index on the angle $\theta$.
The analysis technique employed by Tessein et al. (2009) was based on 1 hour intervals 
of solar wind data with a uniform cadence of 64 seconds.  The spectral exponent for 
each 1 hour interval was determined
by means of second order structure functions and the mean magnetic field for each interval, 
a 1 hour average, was used to bin the data according to the angle $\theta$.  
\medskip

Tessein et al. (2009) may have been
unable to detect any angle dependence of the spectral index, in part, because they
neglected the scale dependence of the local mean magnetic field.  This is important 
because the dynamically relevant mean magnetic field for fluctuations at the 10 second 
scale is not the mean magnetic field at the 1 hour scale.  Note, however, that even if 
they did use a scale dependent mean magnetic field, the 1 hour intervals they use probably 
contain insufficient data to measure the spectral index near the parallel direction 
$\theta=0$, at least for the high-speed streams studied here.  To demonstrate this,
the wavelet analysis technique was applied to data intervals with a duration of 2 hours,
subintervals of the data listed in Table 1.  The 2 hour intervals are comparable in length
to the 1 hour intervals used by Tessein et al. (2009).  It was found that the two or three 
angle bins closest to $\theta=0$ usually had insufficient data to form a power spectrum.  
Therefore, when restricted to such short record lengths, the wavelet technique is unable to 
detect changes in the spectral exponent near $\theta=0$.  Even with 8 Hz data, record lengths
of a few hours do not contain sufficient data near the parallel direction to
reveal the change in spectral index.  Like the wavelet analysis,
the analysis based on second order structure functions  
should be able to detect the angle dependence of the spectral index provided a large enough
statistical sample is used and provided the scale dependence of the local 
mean magnetic field is taken into account.  
\medskip

There appears to be a discrepancy between the work reported here and the work of 
\citet{Dasso:2005}.  \citet{Dasso:2005} used data from the Advanced Composition Explorer (ACE)
in the ecliptic plane near 1 AU to study correlation functions of solar wind 
fluctuations as functions of the angle $\theta$ and concluded that in
high-speed streams the power in the fluctuations is dominated by wavevectors nearly parallel to the
local mean magnetic field.  This result of Dasso et al. (2005) appears to be inconsistent with the
results presented here which show that for each of the
high-speed streams investigated $P_\perp/P_\parallel>1$ throughout the inertial range, that is,
the power when the spacecraft traverses a path perpendicular to $\bm B_0$ is greater than
the power when the spacecraft traverses a path parallel to $\bm B_0$.
This implies that power in the 3D wavevector spectrum is dominated by fluctuations
with $k_\perp\gg k_\parallel$.  This is also a well known feature of MHD turbulence
with a strong mean magnetic field \citep{Shebalin:1983, Oughton:1994, Matthaeus_Ghosh:1996}.
It should be noted that the study of \citet{Dasso:2005} encompasses larger inertial 
range scales than those considered here.  However, it is not known if this can
explain the discrepancy.
\medskip

\citet{Hamilton:2008} studied the variance anisotropy of magnetic field fluctuations
$(\delta B_\perp)^2/(\delta B_\parallel)^2$, the ratio  of the power in the two 
{\it components} of the magnetic field vector perpendicular to $\bm B_0$ to the 
component parallel to $\bm B_0$.  Their results show that for $\beta_p\sim 1$ the ratio
$(\delta B_\perp/\delta B_\parallel)^2$ takes typical values between 1 and 5,
both in the inertial range and dissipation range.  Moreover, in the dissipation 
range this ratio is typically less than or equal to that in the inertial range 
indicating an increase in the parallel component relative to the perpendicular
components. 
This behavior is qualitatively consistent with theoretical predictions for KAWs
(Hollweg 1999).
The power ratio $P_\perp/P_\parallel$ in the present study is different 
from the variance anisotropy investigated by \citet{Hamilton:2008} because $P_\perp$
and $P_\parallel$ both represent the trace power (three vector components).  The wavelet 
technique can easily be adapted to measure the
powers of the magnetic field components and the associated variance anisotropy.
This is an interesting avenue of investigation for future work.

\subsection{Theoretical interpretation}

The observed variations of the spectral exponent as a function of $\theta$  
have a theoretical interpretation that sheds light on the 3D wavevector anisotropy of solar
wind turbulence in high speed wind.  Horbury et al. (2008) have suggested the 
results may be interpreted as evidence that solar wind turbulence is 
characterized by a 3D wavevector spectrum
of the form given in the Goldreich \& Sridhar (1995) theory of strong incompressible MHD turbulence,
a theory based in part on the fundamental work by \citet{Higdon:1984}.  This is because
the functional form of the Goldreich-Sridhar
spectrum implies that the reduced spectra, the spectra observed in the spacecraft frame, have
power-law exponents of 5/3 and 2 in the perpendicular and parallel directions, respectively;
see Tessein et al. (2009) for a derivation of this result.  However,
it should be noted that Boldyrev's (2006) theory of strong incompressible MHD turbulence 
gives a similar prediction with power-law exponents of 3/2 and 2 in the perpendicular and 
parallel directions, respectively; the appropriate form of the wavevector spectrum in 
this case is discussed by \citet{Podesta:2009}. The analysis of solar wind data shows that
the perpendicular power-law exponents 3/2 and 5/3 cannot be distinguished from the data 
because of experimental uncertainties.  Therefore, it may be premature to draw inferences 
regarding agreement with one or another theory.
\medskip

Simulations of incompressible MHD turbulence suggest that Boldyrev's scaling applies
in the presence of a strong mean magnetic field such that $\delta b \ll B_0$, where
$\delta b$ is the r.m.s.\ level of the turbulence and $B_0$ is the magnitude of the mean
magnetic field \citep{Mason:2006,Mason:2008}.  These and other simulations also suggest that the 
scaling may revert to the Goldreich-Sridhar scaling when the mean magnetic field is weak,
$\delta b \gtrsim B_0$, although Boldyrev (2006) has speculated that his scaling may
still apply in the inertial range provided the inequality $\delta b \ll B_0$ holds
for fluctuation amplitudes $\delta b$ within the inertial range.  For both the Ulysses data
analyzed by Horbury et al. (2008) and the high-speed streams analyzed here, the 
r.m.s.\ level of the turbulence satisfies $\delta b \sim B_0$ while at the smallest inertial
range scales $\delta b \ll B_0$.  Observations of the 
power ratio $P_\perp/P_\parallel$ in Figure \ref{Ulysses} show that for the Ulysses data
the Goldreich-Sridhar scaling gives a better fit than the Boldyrev scaling.  Therefore, 
observations of this one time interval indicate that the Goldreich-Sridhar scaling is more
consistent with the data even though $\delta b \ll B_0$ at the smallest inertial
range scales.  The scaling law of the power ratio $P_\perp/P_\parallel$ for high-speed streams
in the ecliptic plane
is difficult to determine because there is insufficient data to obtain reliable measurements 
at low frequencies.  This problem can be resolved by using longer record lengths, however, the
record length is limited by the lifetime of the streams.
\medskip

From a theoretical point of view, it is important to realize that neither of the above theories 
can be applied to solar wind turbulence, even if incompressibility is a reasonable first approximation
for the description of large MHD-scale solar wind fluctuations.  This is
because the theories of Goldreich \& Sridhar (1995) and Boldyrev (2006)
both assume that the cross-helicity of the turbulence vanishes, a point also made explicit by 
Higdon (1984), an assumption that is usually not satisfied in the 
solar wind \citep{Marsch_Tu:1990, Marsch:1991}.  A generalization of these theories to
turbulence with nonvanishing cross-helicity is required to provide the proper foundation
for comparisons between turbulence theory and solar wind observations.  In particular, 
the theoretical form of the 3D wavevector spectrum of the two Elsasser spectra $E^+$ and $E^-$
is needed.
\medskip

Lithwick et al. (2007) have extended the theory of Goldreich \& Sridhar (1995) to imbalanced 
turbulence, that is, turbulence with nonvanishing cross-helicity.  
In the theory of Lithwick et al. (2007) the parallel correlation lengths of plus and 
minus Elsasser wavepackets are equal, that is, $\lambda_\parallel^+=\lambda_\parallel^-$. 
However, this is inconsistent with the existence of different cascade times for the plus 
and minus Elsasser wavepackets as found in the theory. This inconsistency is easily remedied 
and the perpendicular energy spectra of the two Elsasser fields still have the same 
$k_\perp^{-5/3}$ scaling found by Lithwick et al. (2007).  With this modification, the 
energy cascades of the plus and minus 
Elsasser variables are both in a state of critical balance and the reduced spectra
perpendicular and parallel to the mean magnetic field are predicted to scale like
$k_\perp^{-5/3}$ and $k_\parallel^{-2}$, respectively.  Likewise, in the 
generalizations of the theory of Boldyrev (2006) to imbalanced turbulence
by \citet{Perez:2009} and \citet{Podesta_Bhattacharjee:2009}, the reduced spectra
perpendicular and parallel to the mean magnetic field are predicted to scale like
$k_\perp^{-3/2}$ and $k_\parallel^{-2}$, respectively. 
The solar wind observations of Horbury et al. (2008) and those presented here 
support these theoretical predictions although they cannot yet distinguish the 
Goldreich \& Sridhar scaling $k_\perp^{-5/3}$ from the 
Boldyrev scaling $k_\perp^{-3/2}$ based on power-law exponents alone.
\medskip

A remarkable conclusion of this study is that
single spacecraft measurements can be used to glean information about the
wavevector anisotropy of solar wind turbulence.  Whether the power spectrum
measured in the direction perpendicular to the local mean magnetic field has an exponent 
of 5/3 or 3/2 is an important fundamental question that remains unanswered. 
Interestingly, the 3/2 spectrum for magnetic field fluctuations is not yet ruled out.  
\medskip

It is a pleasure to thank Amitava Bhattacharjee, Charles Smith, Bernard Vasquez,
Sergei Markovskii, Mario Acu\~na, David Curtis, and Jean Perez for helpful discussions.
I am especially grateful to S. Peter Gary for his interest in this work and for
several helpful comments and suggestions.

\appendix

\section{Derivation of the inversion formula}

\indent\indent 
Equation (\ref{P}) is derived from equation (\ref{Parseval}) as follows.  The Fourier transform
of the continuous wavelet transform (\ref{xfm}) is
\begin{equation}
\int_{-\infty}^\infty F(s,t) e^{-i\omega t} \, dt = |s|^{1/2} \hat f(\omega) \hat\psi^*(\omega s),
\end{equation}
where $\hat f(\omega)$ and $\hat \psi(\omega)$ are the Fourier transforms of $f(t)$ and 
$\psi(t)$, respectively.  Therefore, by Parseval's theorem for Fourier transforms,
\begin{equation}
\int_{-\infty}^\infty \int_{-\infty}^\infty 
F(s,t)G^*(s,t)\, \frac{ds \,dt}{s^2}=\int_{-\infty}^\infty \frac{ds}{|s|}
\int_{-\infty}^\infty \frac{d\omega}{2\pi} \, \hat f(\omega) \hat g^*(\omega) |\hat\psi(\omega s)|^2
\label{A1}
\end{equation}
Assuming $\hat \psi(\omega)=0$ for $\omega<0$, the right-hand side of equation (\ref{A1}) is equal to 
\begin{multline}
\int_{0}^\infty \frac{ds}{|s|}
\int_{0}^\infty \frac{d\omega}{2\pi} \, \hat f(\omega) \hat g^*(\omega) |\hat\psi(\omega s)|^2 
+\int_{-\infty}^0 \frac{ds}{|s|}
\int_{-\infty}^0 \frac{d\omega}{2\pi} \, \hat f(\omega) \hat g^*(\omega) |\hat\psi(\omega s)|^2  \\
= \int_{0}^\infty \frac{ds}{|s|}
\int_{0}^\infty \frac{d\omega}{2\pi} \, [\hat f(\omega) \hat g^*(\omega)+\hat f(-\omega) \hat g^*(-\omega)] 
|\hat\psi(\omega s)|^2. 
\label{A2}
\end{multline}
If the functions $f(t)$ and $g(t)$ are real valued, then $\hat f(-\omega)=\hat f^*(\omega)$ and
this becomes
\begin{equation}
=2\, \mbox{Re}\int_{0}^\infty \frac{ds}{|s|}
\int_{0}^\infty \frac{d\omega}{2\pi} \, \hat f(\omega) \hat g^*(\omega) |\hat\psi(\omega s)|^2.
\end{equation}
Changing the order of integration and performing the integral with respect to $s$ yields
\begin{equation}
=2C\, \mbox{Re}
\int_{0}^\infty \frac{d\omega}{2\pi} \, \hat f(\omega) \hat g^*(\omega) 
=C \int_{-\infty}^\infty f(t)g(t)\, dt
\end{equation}
(write the integral on the left-hand side as an integral from $-\infty$ to $+\infty$
and then apply Parseval's theorem).  Now, from (\ref{A2}), this is also equal to
\begin{equation}
2\, \mbox{Re}\int_{0}^\infty \frac{ds}{|s|}
\int_{-\infty}^\infty \frac{d\omega}{2\pi} \, \hat f(\omega) \hat g^*(\omega) |\hat\psi(\omega s)|^2
=2\, \mbox{Re}\int_{0}^\infty \frac{ds}{s^2}
\int_{-\infty}^\infty dt\; F(s,t)G^*(s,t).
\end{equation}
Hence, 
\begin{equation}
2\, \mbox{Re}\int_{0}^\infty \frac{ds}{s^2}
\int_{-\infty}^\infty dt\; F(s,t)G^*(s,t)=C \int_{-\infty}^\infty f(t)g(t)\, dt.
\end{equation}
Setting $g(t)=\delta(t-\tau)$ yields equation (\ref{inv}) and setting $g=f$ yields
equation (\ref{P}).

\section{Wavelet transform of $\cos(\omega t)$}

\indent\indent 
Inserting $f(t)=\cos(\omega t)$ into equation (\ref{xfm}) yields the continuous
wavelet transform
\begin{equation}
F(s,t)=  2^{-1/2}\pi^{1/4} |s|^{1/2} [A\exp(i\omega t)+B\exp(-i\omega t)],
\end{equation}
where
\begin{align}
A &=  \exp[-(\omega s -\omega_0)^2/2]-\exp\{-[(\omega s)^2 +\omega_0^2]/2\}, \\
B &=  \exp[-(\omega s +\omega_0)^2/2]-\exp\{-[(\omega s)^2 +\omega_0^2]/2\}.
\end{align}
The distribution of energy in the $(s,t)$ plane is given by the function
\begin{equation}
|F(s,t)|^2=  \frac{\pi^{1/2}}{2} |s|(A^2+B^2) \left[1+\frac{2AB}{A^2+B^2}\cos(2\omega t)\right].
\end{equation}
It is easy to show that $A/B=-\exp(\omega \omega_0 s)$ and, therefore, 
\begin{equation}
|F(s,t)|^2=  \frac{\pi^{1/2}}{2} |s|(A^2+B^2) \left[1-\frac{\cos(2\omega t)}{\cosh(\omega \omega_0 s)}\right].
\end{equation}
The last term in the square brackets is negligible when $\omega \omega_0 s\ge 3$, that is,
it is negligible except at very small frequencies or small scales.  With this approximation,
\begin{equation}
|F(s,t)|^2 \simeq  \frac{\pi^{1/2}}{2} |s|(A^2+B^2).
\end{equation}
To find the scale $s$ where this function attains its maximum, assume $s>0$ and then differentiate this 
expression with respect to $s$ and set the result equal to zero.  This yields the equation 
\begin{equation}
\phi(x)= \bigg(\frac{2x^2}{\omega_0^2}-1\bigg) \cosh(x)-x\frac{\cosh(3x/2)}{\sinh(x/2)}=0,
\end{equation}
where $x=\omega \omega_0 s$.  This equation has exactly one root in the interval $0<x<\infty$.
It has already been assumed that $\cosh(x)\gg 1$ and, therefore, this may be approximated by
the quadratic equation  
\begin{equation}
\frac{2x^2}{\omega_0^2}-2x-1=0
\end{equation}
with solution $x\simeq \omega_0^2+\frac{1}{2}$.  Hence, the scale $s$ where the wavelet 
transform of $\cos(\omega t)$ is a maximum is approximately
\begin{equation}
s= \frac{1}{\omega \omega_0} \big(\omega_0^2+{\textstyle \frac{1}{2}}\big).
\end{equation}
This proves (\ref{nu_x}).

\bibliographystyle{newapa}
\bibliography{jp}

\hyphenation{Post-Script Sprin-ger}
\begin{thebibliography}{}

\bibitem[\protect\citeauthoryear{{Acu{\~n}a}, {Curtis}, {Scheifele}, {Russell},
  {Schroeder}, {Szabo} \& {Luhmann}}{{Acu{\~n}a} et~al.}{2008}]{Acuna:2008}
{Acu{\~n}a}, M.~H., {Curtis}, D., {Scheifele}, J.~L., {Russell}, C.~T.,
  {Schroeder}, P., {Szabo}, A., \& {Luhmann}, J.~G. (2008).
\newblock {The STEREO/IMPACT Magnetic Field Experiment}.
\newblock {\em Space Sci.\ Rev.}, {\em 136}, 203--226.

\bibitem[\protect\citeauthoryear{{Bale}, {Kellogg}, {Mozer}, {Horbury} \&
  {Reme}}{{Bale} et~al.}{2005}]{Bale:2005}
{Bale}, S.~D., {Kellogg}, P.~J., {Mozer}, F.~S., {Horbury}, T.~S., \& {Reme},
  H. (2005).
\newblock {Measurement of the Electric Fluctuation Spectrum of
  Magnetohydrodynamic Turbulence}.
\newblock {\em Phys.\ Rev.\ Lett.}, {\em 94\/}(21), 215002.

\bibitem[\protect\citeauthoryear{{Balogh}, {Beek}, {Forsyth}, {Hedgecock},
  {Marquedant}, {Smith}, {Southwood} \& {Tsurutani}}{{Balogh}
  et~al.}{1992}]{Balogh:1992}
{Balogh}, A., {Beek}, T.~J., {Forsyth}, R.~J., {Hedgecock}, P.~C.,
  {Marquedant}, R.~J., {Smith}, E.~J., {Southwood}, D.~J., \& {Tsurutani},
  B.~T. (1992).
\newblock {The magnetic field investigation on the ULYSSES mission -
  Instrumentation and preliminary scientific results}.
\newblock {\em Astron.\ Astrophys.\ Suppl.\ Ser.}, {\em 92}, 221--236.

\bibitem[\protect\citeauthoryear{{Behannon}}{{Behannon}}{1976}]{Behannon:1976}
{Behannon}, K.~W. (1976).
\newblock {\em {Observations of the interplanetary magnetic field between 0.46
  and 1 A.U. by the Mariner 10 spacecraft}}.
\newblock PhD thesis, AA(National Aeronautics and Space Administration.~Goddard
  Space Flight Center, Greenbelt, MD.).

\bibitem[\protect\citeauthoryear{{Beresnyak} \& {Lazarian}}{{Beresnyak} \&
  {Lazarian}}{2008}]{Beresnyak_Lazarian:2008}
{Beresnyak}, A. \& {Lazarian}, A. (2008).
\newblock {Strong Imbalanced Turbulence}.
\newblock {\em Astrophys.\ J.}, {\em 682}, 1070--1075.

\bibitem[\protect\citeauthoryear{{Bieber}, {Wanner} \& {Matthaeus}}{{Bieber}
  et~al.}{1996}]{Bieber:1996}
{Bieber}, J.~W., {Wanner}, W., \& {Matthaeus}, W.~H. (1996).
\newblock {Dominant two-dimensional solar wind turbulence with implications for
  cosmic ray transport}.
\newblock {\em J.\ Geophys.\ Res.}, {\em 101}, 2511--2522.

\bibitem[\protect\citeauthoryear{{Boldyrev}}{{Boldyrev}}{2005}]{Boldyrev:2005}
{Boldyrev}, S. (2005).
\newblock {On the Spectrum of Magnetohydrodynamic Turbulence}.
\newblock {\em Astrophys.\ J.\ Lett.}, {\em 626}, L37--L40.

\bibitem[\protect\citeauthoryear{{Boldyrev}}{{Boldyrev}}{2006}]{Boldyrev:2006}
{Boldyrev}, S. (2006).
\newblock {Spectrum of Magnetohydrodynamic Turbulence}.
\newblock {\em Phys. Rev. Lett.}, {\em 96\/}(11), 115002.

\bibitem[\protect\citeauthoryear{{Briggs} \& {Henson}}{{Briggs} \&
  {Henson}}{1995}]{Briggs_Henson:1995}
{Briggs}, W.~L. \& {Henson}, V.~E. (1995).
\newblock {\em {The DFT: An Owners Manual for the Discrete Fourier Transform}}.
\newblock SIAM.

\bibitem[\protect\citeauthoryear{{Chandran}}{{Chandran}}{2008}]{Chandran:2008}
{Chandran}, B.~D.~G. (2008).
\newblock {Strong Anisotropic MHD Turbulence with Cross Helicity}.
\newblock {\em Astrophs.\ J.}, {\em 685}, 646--658.

\bibitem[\protect\citeauthoryear{{Cranmer} \& {van Ballegooijen}}{{Cranmer} \&
  {van Ballegooijen}}{2003}]{Cranmer:2003}
{Cranmer}, S.~R. \& {van Ballegooijen}, A.~A. (2003).
\newblock {Alfv{\'e}nic Turbulence in the Extended Solar Corona: Kinetic
  Effects and Proton Heating}.
\newblock {\em Astrophys.\ J.}, {\em 594}, 573--591.

\bibitem[\protect\citeauthoryear{{Dasso}, {Milano}, {Matthaeus} \&
  {Smith}}{{Dasso} et~al.}{2005}]{Dasso:2005}
{Dasso}, S., {Milano}, L.~J., {Matthaeus}, W.~H., \& {Smith}, C.~W. (2005).
\newblock {Anisotropy in Fast and Slow Solar Wind Fluctuations}.
\newblock {\em Astrophys. J.}, {\em 635}, L181--L184.

\bibitem[\protect\citeauthoryear{{Daubechies}}{{Daubechies}}{1992}]{Daubechies%
:1992}
{Daubechies}, I. (1992).
\newblock {\em {Ten lectures on wavelets }}.
\newblock Society for Industrial and Applied Mathematics.

\bibitem[\protect\citeauthoryear{{Galvin}, {Kistler}, {Popecki}, {Farrugia},
  {Simunac}, {Ellis}, {M{\"o}bius}, {Lee}, {Boehm}, {Carroll}, {Crawshaw},
  {Conti}, {Demaine} \& {Ellis}}{{Galvin} et~al.}{2008}]{Galvin:2008}
{Galvin}, A.~B., {Kistler}, L.~M., {Popecki}, M.~A., {Farrugia}, C.~J.,
  {Simunac}, K.~D.~C., {Ellis}, L., {M{\"o}bius}, E., {Lee}, M.~A., {Boehm},
  M., {Carroll}, J., {Crawshaw}, A., {Conti}, M., {Demaine}, P., \& {Ellis}, S.
  (2008).
\newblock {The Plasma and Suprathermal Ion Composition (PLASTIC) Investigation
  on the STEREO Observatories}.
\newblock {\em Space Sci.\ Rev.}, {\em 136}, 437--486.

\bibitem[\protect\citeauthoryear{{Gautschi}}{{Gautschi}}{1970}]{Gautschi:1970}
{Gautschi}, W. (1970).
\newblock {Efficient computation of the complex error function}.
\newblock {\em SIAM J.\ Numer. Anal.}, {\em 7}, 187--198.

\bibitem[\protect\citeauthoryear{{Goldreich} \& {Sridhar}}{{Goldreich} \&
  {Sridhar}}{1995}]{Goldreich_Sridhar:1995}
{Goldreich}, P. \& {Sridhar}, S. (1995).
\newblock {Toward a theory of interstellar turbulence. 2: Strong alfvenic
  turbulence}.
\newblock {\em Astrophys. J.}, {\em 438}, 763--775.

\bibitem[\protect\citeauthoryear{{Goldreich} \& {Sridhar}}{{Goldreich} \&
  {Sridhar}}{1997}]{Goldreich_Sridhar:1997}
{Goldreich}, P. \& {Sridhar}, S. (1997).
\newblock {Magnetohydrodynamic Turbulence Revisited}.
\newblock {\em Astrophys. J.}, {\em 485}, 680.

\bibitem[\protect\citeauthoryear{{Hamilton}, {Smith}, {Vasquez} \&
  {Leamon}}{{Hamilton} et~al.}{2008}]{Hamilton:2008}
{Hamilton}, K., {Smith}, C.~W., {Vasquez}, B.~J., \& {Leamon}, R.~J. (2008).
\newblock {Anisotropies and helicities in the solar wind inertial and
  dissipation ranges at 1 AU}.
\newblock {\em Journal of Geophysical Research (Space Physics)}, {\em
  113\/}(12), 1106.

\bibitem[\protect\citeauthoryear{{Higdon}}{{Higdon}}{1984}]{Higdon:1984}
{Higdon}, J.~C. (1984).
\newblock {Density fluctuations in the interstellar medium: Evidence for
  anisotropic magnetogasdynamic turbulence. I - Model and astrophysical sites}.
\newblock {\em Astroph.\ J.}, {\em 285}, 109--123.

\bibitem[\protect\citeauthoryear{{Hollweg}}{{Hollweg}}{1999}]{Hollweg:1999}
{Hollweg}, J.~V. (1999).
\newblock {Kinetic Alfv{\'e}n wave revisited}.
\newblock {\em J.\ Geophys.\ Res.}, {\em 104}, 14811--14820.

\bibitem[\protect\citeauthoryear{{Horbury}, {Forman} \& {Oughton}}{{Horbury}
  et~al.}{2008}]{Horbury_Forman:2008}
{Horbury}, T.~S., {Forman}, M., \& {Oughton}, S. (2008).
\newblock {Anisotropic Scaling of Magnetohydrodynamic Turbulence}.
\newblock {\em Phys.\ Rev.\ Lett.}, {\em 101\/}(17), 175005.

\bibitem[\protect\citeauthoryear{{Howes}}{{Howes}}{2008}]{Howes:2008a}
{Howes}, G.~G. (2008).
\newblock {Inertial range turbulence in kinetic plasmas}.
\newblock {\em Phys.\ Plasmas}, {\em 15\/}(5), 055904.

\bibitem[\protect\citeauthoryear{{Howes}, {Cowley}, {Dorland}, {Hammett},
  {Quataert} \& {Schekochihin}}{{Howes} et~al.}{2008}]{Howes:2008b}
{Howes}, G.~G., {Cowley}, S.~C., {Dorland}, W., {Hammett}, G.~W., {Quataert},
  E., \& {Schekochihin}, A.~A. (2008).
\newblock {A model of turbulence in magnetized plasmas: Implications for the
  dissipation range in the solar wind}.
\newblock {\em J.\ Geophys.\ Res.}, {\em 113\/}(12), 5103.

\bibitem[\protect\citeauthoryear{{Jian}, {Russell}, {Luhmann}, {Blanco-Cano},
  {Leisner}, {Galvin}, {Neubauer} \& {Wennmacher}}{{Jian}
  et~al.}{2008}]{Jian:2008}
{Jian}, L.~K., {Russell}, C.~T., {Luhmann}, J.~G., {Blanco-Cano}, X.,
  {Leisner}, J.~S., {Galvin}, A.~B., {Neubauer}, F.~M., \& {Wennmacher}, A.
  (2008).
\newblock {Ion Cyclotron Waves in the Solar Wind at 0.3 and 1 AU}.
\newblock {\em AGU Fall Meeting Abstracts}, {\em 0}, 1574.

\bibitem[\protect\citeauthoryear{{Leamon}, {Ness} \& {Smith}}{{Leamon}
  et~al.}{1999}]{Leamon:1999b}
{Leamon}, R., {Ness}, N.~F., \& {Smith}, C.~W. (1999).
\newblock {The Dynamics of Dissipation Range Fluctuations with Application to
  Cosmic Ray Propagation Theory}.
\newblock In {Kieda}, D., {Salamon}, M., \& {Dingus}, B. (Eds.), {\em
  Proceedings of the 26th International Cosmic Ray Conference}, volume~6 of
  {\em International Cosmic Ray Conference, August 17-25, 1999. Salt Lake City,
  Utah}, (pp.\ 366--369).

\bibitem[\protect\citeauthoryear{{Leamon}, {Smith}, {Ness} \& {Wong}}{{Leamon}
  et~al.}{1999}]{Leamon:1999c}
{Leamon}, R.~J., {Smith}, C.~W., {Ness}, N.~F., \& {Wong}, H.~K. (1999).
\newblock {Dissipation range dynamics: Kinetic Alfv{\'e}n waves and the
  importance of {$\beta_e$}}.
\newblock {\em J.\ Geophys.\ Res.}, {\em 104}, 22331--22344.

\bibitem[\protect\citeauthoryear{{Lithwick}, {Goldreich} \&
  {Sridhar}}{{Lithwick} et~al.}{2007}]{Lithwick_Goldreich:2007}
{Lithwick}, Y., {Goldreich}, P., \& {Sridhar}, S. (2007).
\newblock {Imbalanced Strong MHD Turbulence}.
\newblock {\em Astrophs.\ J.}, {\em 655}, 269--274.

\bibitem[\protect\citeauthoryear{{Luhmann}, {Curtis}, {Schroeder}, {McCauley},
  {Lin}, {Larson}, {Bale}, {Sauvaud}, {Aoustin}, {Mewaldt}, {Cummings}, {Stone}
  \& {Davis}}{{Luhmann} et~al.}{2008}]{Luhmann:2008}
{Luhmann}, J.~G., {Curtis}, D.~W., {Schroeder}, P., {McCauley}, J., {Lin},
  R.~P., {Larson}, D.~E., {Bale}, S.~D., {Sauvaud}, J.-A., {Aoustin}, C.,
  {Mewaldt}, R.~A., {Cummings}, A.~C., {Stone}, E.~C., \& {Davis}, A.~J.
  (2008).
\newblock {STEREO IMPACT Investigation Goals, Measurements, and Data Products
  Overview}.
\newblock {\em Space Sci.\ Rev.}, {\em 136}, 117--184.

\bibitem[\protect\citeauthoryear{{Marsch}}{{Marsch}}{1991a}]{Marsch:1991b}
{Marsch}, E. (1991a).
\newblock {\em {Kinetic Physics of the Solar Wind Plasma}}, (pp.\ 45--133).
\newblock Physics of the Inner Heliosphere II.~ Particles, Waves and
  Turbulence, Springer-Verlag Berlin.~ Also Physics and Chemistry in Space,
  volume 21; 2.

\bibitem[\protect\citeauthoryear{{Marsch}}{{Marsch}}{1991b}]{Marsch:1991}
{Marsch}, E. (1991b).
\newblock {\em {MHD Turbulence in the Solar Wind}}, (pp.\ 159--241).
\newblock Physics of the Inner Heliosphere II.~ Particles, Waves and
  Turbulence, Springer-Verlag Berlin.~ Also Physics and Chemistry in Space,
  volume 21; 2.

\bibitem[\protect\citeauthoryear{{Marsch}}{{Marsch}}{2006}]{Marsch:2006}
{Marsch}, E. (2006).
\newblock {Kinetic Physics of the Solar Corona and Solar Wind}.
\newblock {\em Living Reviews in Solar Physics}, {\em 3}, 1.

\bibitem[\protect\citeauthoryear{{Marsch} \& {Tu}}{{Marsch} \&
  {Tu}}{1990}]{Marsch_Tu:1990}
{Marsch}, E. \& {Tu}, C.-Y. (1990).
\newblock {On the radial evolution of MHD turbulence in the inner heliosphere}.
\newblock {\em J.\ Geophys.\ Res.}, {\em 95}, 8211--8229.

\bibitem[\protect\citeauthoryear{{Mason}, {Cattaneo} \& {Boldyrev}}{{Mason}
  et~al.}{2006}]{Mason:2006}
{Mason}, J., {Cattaneo}, F., \& {Boldyrev}, S. (2006).
\newblock {Dynamic Alignment in Driven Magnetohydrodynamic Turbulence}.
\newblock {\em Phys.\ Rev.\ Lett.}, {\em 97\/}(25), 255002.

\bibitem[\protect\citeauthoryear{{Mason}, {Cattaneo} \& {Boldyrev}}{{Mason}
  et~al.}{2008}]{Mason:2008}
{Mason}, J., {Cattaneo}, F., \& {Boldyrev}, S. (2008).
\newblock {Numerical measurements of the spectrum in magnetohydrodynamic
  turbulence}.
\newblock {\em Phys. Rev. E.}, {\em 77\/}(3), 036403.

\bibitem[\protect\citeauthoryear{{Matthaeus}, {Ghosh}, {Oughton} \&
  {Roberts}}{{Matthaeus} et~al.}{1996}]{Matthaeus_Ghosh:1996}
{Matthaeus}, W.~H., {Ghosh}, S., {Oughton}, S., \& {Roberts}, D.~A. (1996).
\newblock {Anisotropic three-dimensional MHD turbulence}.
\newblock {\em J.\ Geophys.\ Res.}, {\em 101}, 7619--7630.

\bibitem[\protect\citeauthoryear{{Matthaeus} \& {Goldstein}}{{Matthaeus} \&
  {Goldstein}}{1982}]{Matthaeus_Goldstein:1982}
{Matthaeus}, W.~H. \& {Goldstein}, M.~L. (1982).
\newblock {Measurement of the rugged invariants of magnetohydrodynamic
  turbulence in the solar wind}.
\newblock {\em J.\ Geophys.\ Res.}, {\em 87}, 6011--6028.

\bibitem[\protect\citeauthoryear{{Montgomery}, {Gary}, {Feldman} \&
  {Forslund}}{{Montgomery} et~al.}{1976}]{Montgomery:1976}
{Montgomery}, M.~D., {Gary}, S.~P., {Feldman}, W.~C., \& {Forslund}, D.~W.
  (1976).
\newblock {Electromagnetic instabilities driven by unequal proton beams in the
  solar wind}.
\newblock {\em J.\ Geophys.\ Res.}, {\em 81}, 2743--2749.

\bibitem[\protect\citeauthoryear{{Neugebauer}}{{Neugebauer}}{1975}]{Neugebauer%
:1975}
{Neugebauer}, M. (1975).
\newblock {The enhancement of solar wind fluctuations at the proton thermal
  gyroradius}.
\newblock {\em J.\ Geophys.\ Res.}, {\em 80}, 998--1002.

\bibitem[\protect\citeauthoryear{{Neugebauer}}{{Neugebauer}}{1976}]{Neugebauer%
:1976}
{Neugebauer}, M. (1976).
\newblock {Corrections to and comments on the paper 'The enhancement of solar
  wind fluctuations at the proton thermal gyroradius'}.
\newblock {\em J.\ Geophys.\ Res.}, {\em 81}, 2447.

\bibitem[\protect\citeauthoryear{{Neugebauer}, {Wu} \& {Huba}}{{Neugebauer}
  et~al.}{1978}]{Neugebauer:1978}
{Neugebauer}, M., {Wu}, C.~S., \& {Huba}, J.~D. (1978).
\newblock {Plasma fluctuations in the solar wind}.
\newblock {\em J.\ Geophys.\ Res.}, {\em 83}, 1027--1034.

\bibitem[\protect\citeauthoryear{{Oughton}, {Priest} \& {Matthaeus}}{{Oughton}
  et~al.}{1994}]{Oughton:1994}
{Oughton}, S., {Priest}, E.~R., \& {Matthaeus}, W.~H. (1994).
\newblock {The influence of a mean magnetic field on three-dimensional
  magnetohydrodynamic turbulence}.
\newblock {\em J.\ Fluid Mech.}, {\em 280}, 95--117.

\bibitem[\protect\citeauthoryear{{Percival} \& {Walden}}{{Percival} \&
  {Walden}}{1993}]{Percival_Walden:1993}
{Percival}, D.~B. \& {Walden}, A.~T. (1993).
\newblock {\em {Spectral Analysis for Physical Applications}}.
\newblock Cambridge University Press.

\bibitem[\protect\citeauthoryear{{Perez} \& {Boldyrev}}{{Perez} \&
  {Boldyrev}}{2009}]{Perez:2009}
{Perez}, J.~C. \& {Boldyrev}, S. (2009).
\newblock {The role of cross helicity in magnetohydrodynamic turbulence}.
\newblock {\em Phys.\ Rev.\ Lett.}, {\em 102}, 025003.

\bibitem[\protect\citeauthoryear{{Podesta}}{{Podesta}}{2009}]{Podesta:2009}
{Podesta}, J.~J. (2009).
\newblock {Relationship between the shell-averaged energy spectrum and the
  frequency spectrum measured by a single spacecraft in the solar wind}.
\newblock {\em Astrophs.\ J. (in press)}, {\em 000}, 0.

\bibitem[\protect\citeauthoryear{{Podesta} \& {Bhattacharjee}}{{Podesta} \&
  {Bhattacharjee}}{2009}]{Podesta_Bhattacharjee:2009}
{Podesta}, J.~J. \& {Bhattacharjee}, A. (2009).
\newblock {Theory of incompressible MHD turbulence with cross-helicity}.
\newblock {\em Phys.\ Rev.\ Lett. (submitted)}, {\em 000}, 0.

\bibitem[\protect\citeauthoryear{{Quataert}}{{Quataert}}{1998}]{Quataert:1998}
{Quataert}, E. (1998).
\newblock {Particle Heating by Alfvenic Turbulence in Hot Accretion Flows}.
\newblock {\em Astrophs.\ J.}, {\em 500}, 978.

\bibitem[\protect\citeauthoryear{{Quataert} \& {Gruzinov}}{{Quataert} \&
  {Gruzinov}}{1999}]{Quataert:1999}
{Quataert}, E. \& {Gruzinov}, A. (1999).
\newblock {Turbulence and Particle Heating in Advection-dominated Accretion
  Flows}.
\newblock {\em Astrophs.\ J.}, {\em 520}, 248--255.

\bibitem[\protect\citeauthoryear{{Sari} \& {Valley}}{{Sari} \&
  {Valley}}{1976}]{Sari_Valley:1976}
{Sari}, J.~W. \& {Valley}, G.~C. (1976).
\newblock {Interplanetary magnetic field power spectra - Mean field radial or
  perpendicular to radial}.
\newblock {\em J.\ Geophys.\ Res.}, {\em 81}, 5489--5499.

\bibitem[\protect\citeauthoryear{{Schekochihin}, {Cowley}, {Dorland},
  {Hammett}, {Howes}, {Plunk}, {Quataert} \& {Tatsuno}}{{Schekochihin}
  et~al.}{2008}]{Schekochihin:2008}
{Schekochihin}, A.~A., {Cowley}, S.~C., {Dorland}, W., {Hammett}, G.~W.,
  {Howes}, G.~G., {Plunk}, G.~G., {Quataert}, E., \& {Tatsuno}, T. (2008).
\newblock {Gyrokinetic turbulence: a nonlinear route to dissipation through
  phase space}.
\newblock {\em Plasma Physics and Controlled Fusion}, {\em 50\/}(12), 124024.

\bibitem[\protect\citeauthoryear{{Schekochihin}, {Cowley}, {Dorland},
  {Hammett}, {Howes}, {Quataert} \& {Tatsuno}}{{Schekochihin}
  et~al.}{2007}]{Schekochihin:2009}
{Schekochihin}, A.~A., {Cowley}, S.~C., {Dorland}, W., {Hammett}, G.~W.,
  {Howes}, G.~G., {Quataert}, E., \& {Tatsuno}, T. (2007).
\newblock {Astrophysical gyrokinetics: kinetic and fluid turbulent cascades in
  magnetized weakly collisional plasmas}.
\newblock {\em ArXiv e-prints}, {\em 0\/}(0), 0.

\bibitem[\protect\citeauthoryear{{Schwenn}}{{Schwenn}}{2006}]{Schwenn:2006}
{Schwenn}, R. (2006).
\newblock {Solar Wind Sources and Their Variations Over the Solar Cycle}.
\newblock {\em Space Sci.\ Rev.}, {\em 124}, 51--76.

\bibitem[\protect\citeauthoryear{{Shebalin}, {Matthaeus} \&
  {Montgomery}}{{Shebalin} et~al.}{1983}]{Shebalin:1983}
{Shebalin}, J.~V., {Matthaeus}, W.~H., \& {Montgomery}, D. (1983).
\newblock {Anisotropy in MHD turbulence due to a mean magnetic field}.
\newblock {\em J.\ Plasma Phys.}, {\em 29}, 525--547.

\bibitem[\protect\citeauthoryear{{Smith}, {Vasquez} \& {Hamilton}}{{Smith}
  et~al.}{2006}]{Smith:2006}
{Smith}, C.~W., {Vasquez}, B.~J., \& {Hamilton}, K. (2006).
\newblock {Interplanetary magnetic fluctuation anisotropy in the inertial
  range}.
\newblock {\em J.\ Geophys.\ Res. (Space Physics)}, {\em 111\/}(10), 9111.

\bibitem[\protect\citeauthoryear{{Stix}}{{Stix}}{1992}]{Stix:1992}
{Stix}, T.~H. (1992).
\newblock {\em {Waves in plasmas}}.
\newblock New York: American Institute of Physics.

\bibitem[\protect\citeauthoryear{{Tessein}, {Smith}, {MacBride}, {Matthaeus},
  {Forman} \& {Borovsky}}{{Tessein} et~al.}{2009}]{Tessein:2009}
{Tessein}, J.~A., {Smith}, C.~W., {MacBride}, B.~T., {Matthaeus}, W.~H.,
  {Forman}, M.~A., \& {Borovsky}, J.~E. (2009).
\newblock {Spectral Indices for Multi-Dimensional Interplanetary Turbulence at
  1 AU}.
\newblock {\em Astrophs.\ J.}, {\em 692}, 684--693.

\bibitem[\protect\citeauthoryear{{Zirker}}{{Zirker}}{1977}]{Zirker:1977}
{Zirker}, J.~B. (1977).
\newblock {Coronal holes and high-speed wind streams}.
\newblock {\em Rev.\ Geophys.\ and Space Phys.}, {\em 15}, 257--269.

\end{thebibliography}

\IfFileExists{\jobname.bbl}{}
 {\typeout{}
  \typeout{******************************************}
  \typeout{** Please run "bibtex \jobname" to optain}
  \typeout{** the bibliography and then re-run LaTeX}
  \typeout{** twice to fix the references!}
  \typeout{******************************************}
  \typeout{}
 }

\end{document}